\documentclass[reprint,amsmath,amssymb,aps]{revtex4-2}
\usepackage{amsfonts}
\usepackage{amsmath}
\usepackage{amssymb}
\usepackage{mathrsfs}
\usepackage{braket}
\usepackage{subfigure}
\usepackage{graphicx}
\graphicspath{{fig/}}
\usepackage{epstopdf}
\usepackage{float}
\usepackage{color}
\usepackage{bm}
\usepackage{ulem}
\usepackage{xcolor}
\usepackage{appendix}
\usepackage{overpic}
\usepackage{booktabs}
\usepackage{hyperref}
\usepackage{array}
\usepackage{cancel}

\hypersetup{colorlinks=true, linkcolor=blue, filecolor=blue, urlcolor=blue, citecolor=blue}

\begin{document}
\title{Dynamics of entanglement in non-Hermitian system of nonreciprocal coupling
}
\preprint{CTP-SCU/2023005}
\author{Jiguo Wu$^{1,*}$, Shaojie Yang$^{1,2,*,\dagger}$, Xiong Liu$^1$, Zhibo Jiang$^1$, Zhiyou Zhang$^{1,\dagger}$, Rongchun Ge$^{1,\dagger}$}
\affiliation{$^1$College of Physics, Sichuan University, Chengdu 610064, China}
\affiliation{$^2$Department of Physics, Technical University of Munich, 80333 Munich, Germany}
\affiliation{$^*$The authors contributed equally}
\affiliation{$^\dagger$Corresponding authors: shaojie.yang@tum.de,zhangzhiyou@scu.edu.cn,rcge@scu.edu.cn}

\begin{abstract}
The exploration of the effect of non-Hermitian (NH) in quantum systems is gaining renewed momentum as a result of the recent progress in experiments. It has been indicated that the ability to engineer the orders of exceptional points (EPs) can be fundamental to its applications in quantum control, such as accelerating entanglement generation; however, an initial survey of the interactions between different orders of EPs is still missing. In this work, we try to partially address this issue by employing a non-reciprocally coupled cavity system which can be experimentally realizable with a two-dimensional reconfigurable NH gauged laser array. With deliberate design, our systems display both second and third orders of EPs. The sign of a novel, purely quantum continuous entanglement phase transition is observed, which reveals how multi-EP interactions alter entanglement dynamics across different regimes while maintaining accelerated generation. Our findings reveil the genuinely quantum NH physics of higher-order EPs, which has opened up a new avenue for future investigations of the higher-order degeneracy application and the dynamic phase transitions of quantum systems.

\end{abstract}
\maketitle
\section{Introduction}
Non-Hermitian (NH) dynamics is not rare when one examines a subsystem which is interacting with its environments~\cite{moiseyev1998quantum,plenio1998quantum}, and there is a long history of NH system since the early days of quantum theory. The physical motivation has been the finite life-time of various resonant states such as those in the context of nuclear decay and neutron-nucleus scattering \cite{gamow1928quantentheorie,feshbach1954model}. 
Subsequently, the Lindblad master equation \cite{lindblad1976generators} was developed to depict the dynamics of microscopic NH systems, while the effective NH Hamiltonian is used to characterize the macroscopic mean-field behaviors of NH systems \cite{rotter2009non,hufner1967dwba,feshbach1958unified}.

The recent revival in the exploration of the NH system/theory was initiated with the endeavor of extending the framework of Hermitian systems to fulfill the requirement of a real spectrum. A milestone of these investigations is the discovery that complete real spectra can be achieved for non-Hermitian Hamiltonian systems bearing parity-time ($\mathcal{P}\mathcal{T}$) symmetry~\cite{bender1998real}. As the weight of non-Hermitian component increases, generic complex spectra are recovered in the symmetry broken regime. These two regimes are separated by the well known exceptional points (EPs), and phenomenon is called spontaneous $\mathcal{P}\mathcal{T}$ symmetry breaking. The theoretical exposures have spurred an intensive interest from the experimental side which leads to the contest of exploring new physics of NH systems employing all kinds of systems such as gauged laser arrays~\cite{gao2023two,horiuchi2023non,zhang2020tunable}, metamaterials~\cite{makris2008beam,fleury2015invisible,shao2020non,wu2024wave}, atomic systems~\cite{zhang2016observation,peng2016anti,hang2013pt} and so on. The presence of EPs has strongly twisted the geometry of the abstract Hilbert space: coalescence of the eigenmodes occurs at the singular points. This changes dramatically the metric of space and results in rich and intricate dynamical nature of NH systems. Such phenomena bring new opportunities for quantum controlling and relevant fields, and is a main focus of NH physics. Based on this distinctive feature, the extraordinary sensitivity accompanying EPs has been explored for quantum metrology with the development of a diverse array of high-precision detection schemes including mass sensors \cite{djorwe2019exceptional}, nanoparticle detectors \cite{chen2017exceptional,wiersig2014enhancing,qin2019brillouin}, magnetometer \cite{kononchuk2020orientation,kononchuk2022exceptional} and gyroscope \cite{lai2019observation,ren2017ultrasensitive,mao2020enhanced,hokmabadi2019non,
wang2020petermann,mao2022experimental}, etc.

However, the pure quantum feature of a non-Hermitian system has seldom been explored until very recently~\cite{han2023exceptional,sun2023fractional}. Currently, most of the investigations are about the properties of second order EPs. It is indicated that a fourth order EP can speed up the generation of entanglement~\cite{li2023speeding,yuan2026beating}. These results are mainly based on the NH Hamiltonian system with onsite gain and loss. However, there are many interesting NH systems that are defined by nonreciprocal coupling. Mathematically the nonreciprocal system can be transformed into the cases with onsite gain and loss, but generally a global unitary operation will be needed, as a result it will not keep the dynamics as it was before. On the other hand, acceleration compared to the Hermitian case will usually require an effectively larger Rabi frequency~\cite{bender2007faster}, so it is a nontrivial issue how the comparison to the Hermitian can be done fairly. 

In this work, we investigate the quantum feature of NH Hamiltonian system with nonreciprocal coupling. Inspired by the two-dimensional configuration of NH gauged laser array \cite{gao2023two}, we first present an effective three level system (TLS) with $\mathcal{P}\mathcal{T}$ symmetry which bears both a second order and third order EPs. Then, the dynamics of entanglement and quantum fidelity are used to characterize different regions of the dynamical phase around the EPs. Our numerical results show that the landscape of different regions can be effectively mapped out with both physical quantities. The sign of a continuous entanglement phase transition is observed around the EPs. The phenomenon of speeding up entanglement generation is explored for two coupling TLS. The speeding up is confirmed with numerics, and moreover, the footprint of the interference of different order of EPs is observed from the dynamics of entanglement.

\begin{figure*}[t]
    \begin{center}
        \subfigure[\;$\Omega_1=6\times10^8$, real part of eigenenergy]{\includegraphics[width=4.5cm]{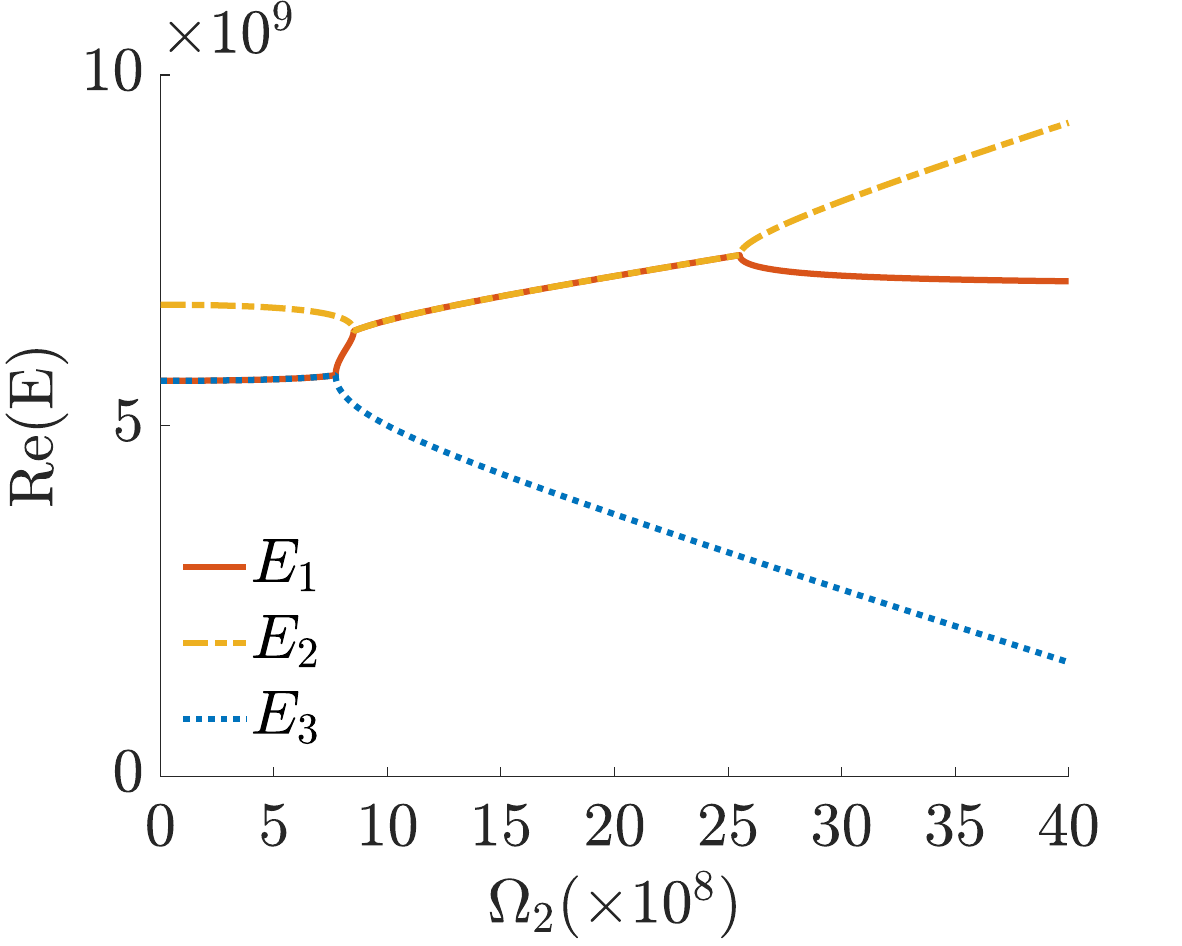}
        \label{6d}}
        \subfigure[\;$\Omega_1=\Omega_p$, real part of eigenenergy]{\includegraphics[width=4.5cm]{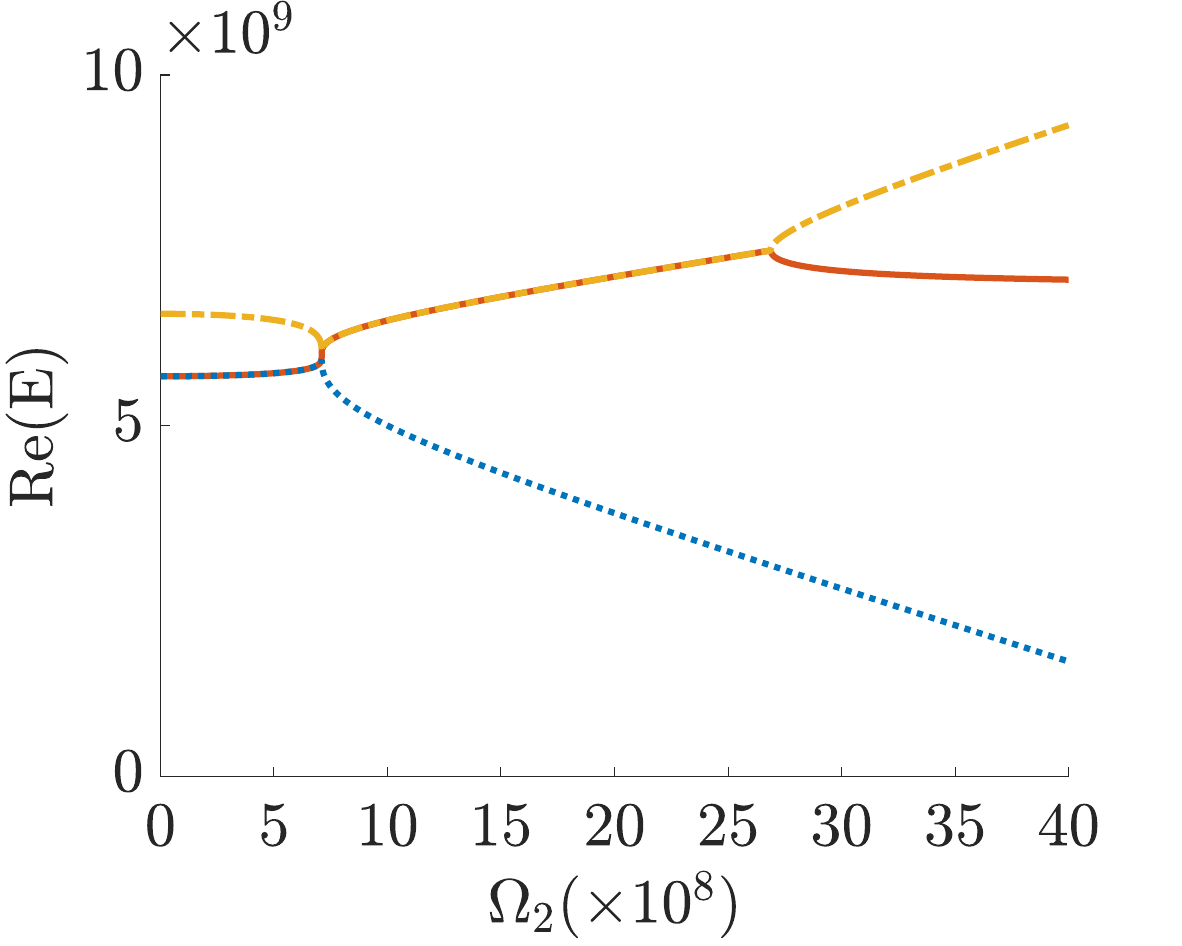}
        \label{6e}}
        \subfigure[\;$\Omega_1=12\times10^8$, real part of eigenenergy]{\includegraphics[width=4.5cm]{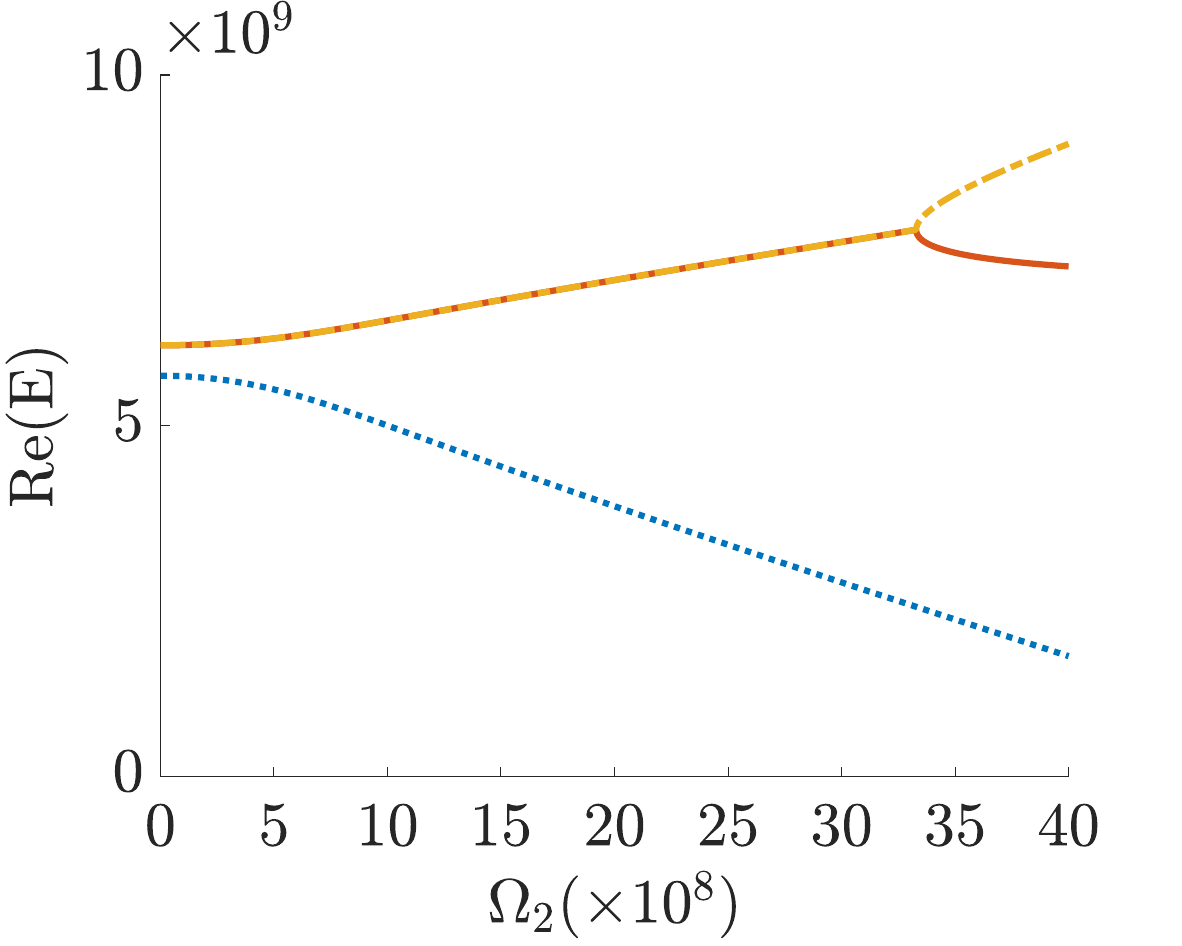}
        \label{6f}}\\
        \subfigure[\;$\Omega_1=6\times10^8$, imaginary part of eigenenergy]{\includegraphics[width=4.5cm]{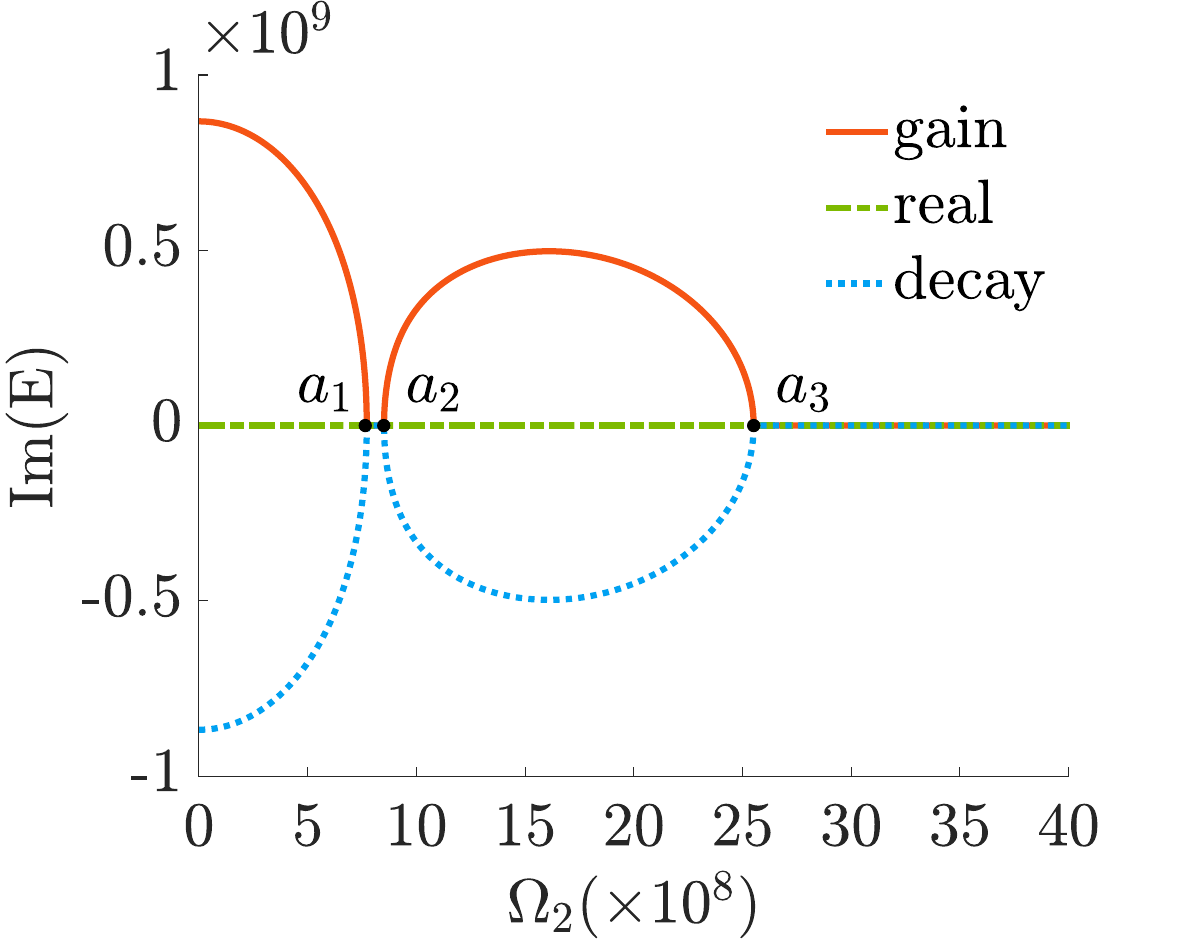}
        \label{6a}}
        \subfigure[\;$\Omega_1=\Omega_p$, imaginary part of eigenenergy]{\includegraphics[width=4.5cm]{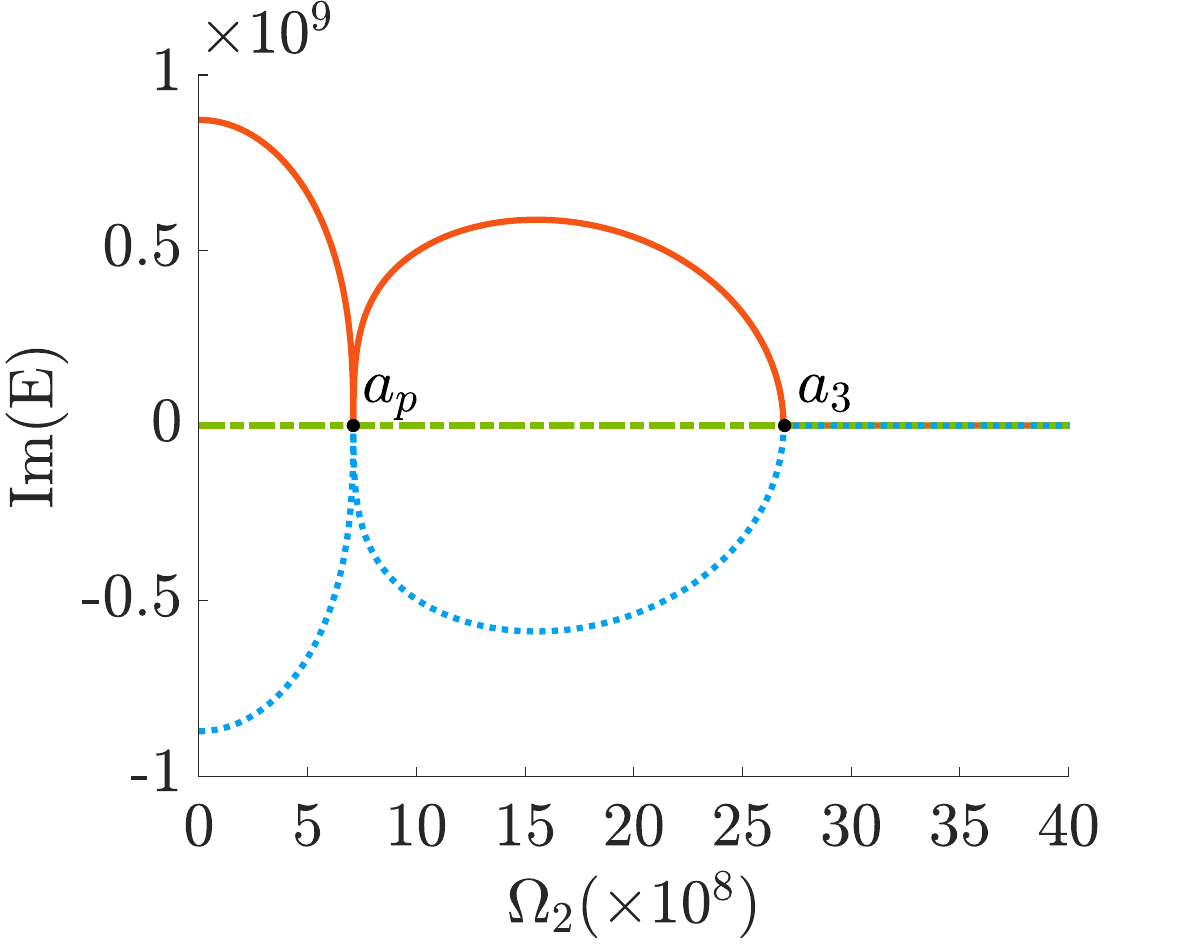}
        \label{6b}}
        \subfigure[\;$\Omega_1=12\times10^8$, imaginary part of eigenenergy]{\includegraphics[width=4.5cm]{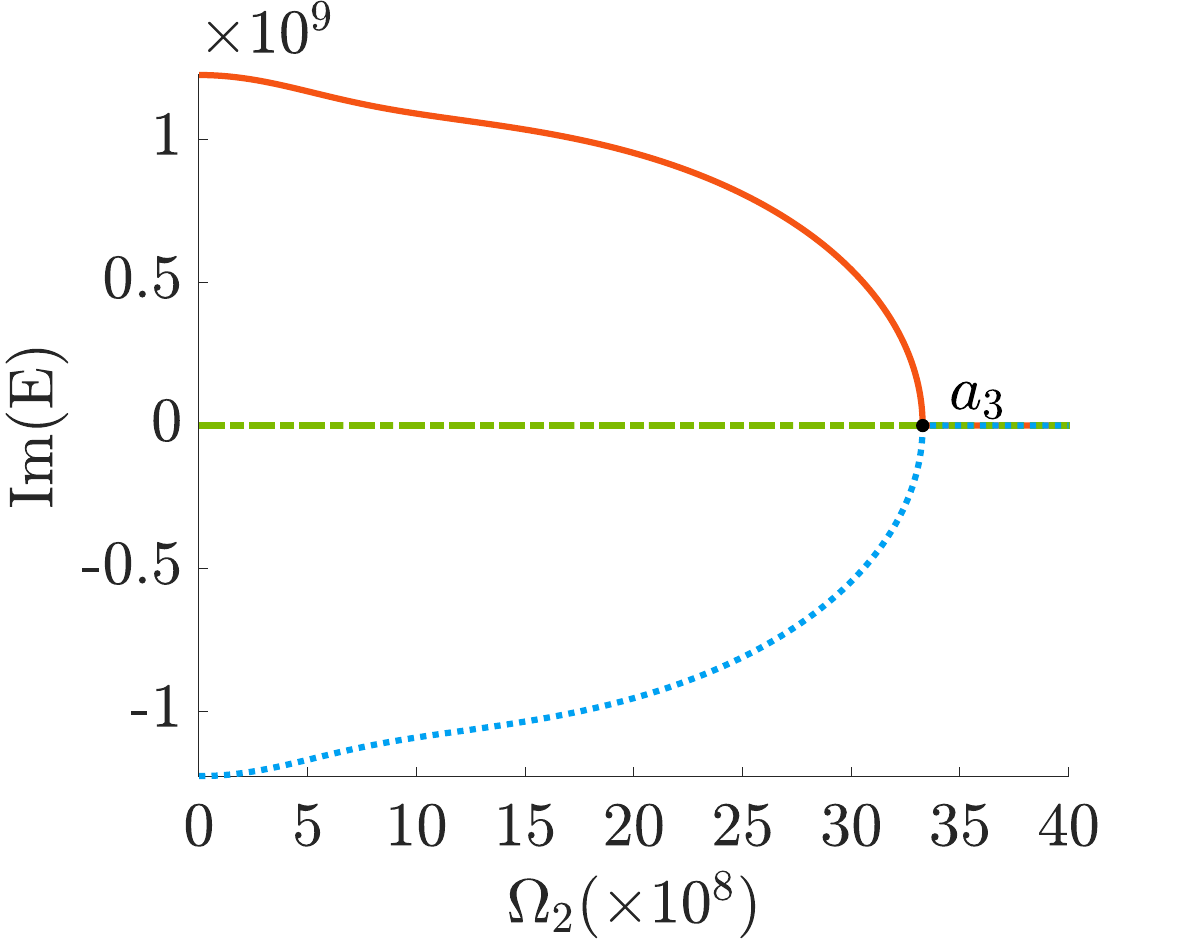}
        \label{6c}}
        \caption{\label{figic}NH energy energy level spectra versus $\Omega_2$ for different value of imaginary coupling. Here we set $w_{A}=5\times10^9$ Hz, $w_{B}=6\times10^9$ Hz, $w_{C}=7\times10^9$ Hz, $e=1\times10^9$ Hz. (a), (d) Real and imaginary parts of the eigenenergies for a small imaginary coupling strength ($\Omega_1=6\times10^8$ Hz). Three $\text{EP}_2$ appear at $\Omega_2=a_1,a_2,a_3$, respectively where pairs of the eigenstates coalesce pairwise. (b),(e)Real and imaginary parts of the eigenenergies at the critical imaginary coupling strength  $\Omega_p$. The $\text{EP}_2$ at $a_1$ and $a_2$ coalesce into a single single $\text{EP}_3$ at $\Omega_p$, which is 3rd-order degenerated. (c),(f) When imaginary coupling is larger than $\Omega_p$, there is no EP at $\Omega_2<a_3$ anymore.}
    \end{center}
\end{figure*}

\section{Model: coupled three-cavity system}
To explore the quantum feature of an NH Hamiltonian system with multiple EPs, we present a minimum model of a three-level quantum system (TLS) which can be idealized as three cavities with nonreciprocal coupling. Cavity A and cavity B are asymmetrically coupled (assuming real tunneling amplitude), while the amplitude among cavity B and cavity C are pure imaginary. Furthermore, in order to obtain a richer topological structure and investigate the interactions between EPs, we break the chiral symmetry by assigning each cavity with a distinct on-site frequency. Under the constraint of photon number conservation, the NH Hamiltonian of our interest is expressed as

\begin{gather}
\begin{split}
\label{eqht}
H&=\sum_iw_ia_i^\dagger a_i+i\Omega_1(a_B^\dagger a_C+a_C^\dagger a_B)\\
&+(\Omega_2-e)a_A^\dagger a_B+(\Omega_2+e)a_B^\dagger a_A,
\end{split}
\end{gather}
where $i \in \{A,B,C\}$ indexes the three cavities, $a^\dagger_i$ and $a_i$ denote the photonic creation and annihilation operators, respectively; real $w_A, w_B, w_C$ are the frequencies of cavity A, cavity B and cavity C, respectively. $\Omega_1$ (real) represents the imaginary coupling strength between cavities B and C, while $(\Omega_2\pm e)$ (real) are the asymmetric coupling strengths between cavities A and B. The photon states of cavity A, B, and C are denoted as $|A\rangle$, $|B\rangle$, $|C\rangle$ respectively. In the following, we concentrate on the sector with total excitation number one. As we plan to investigate the effect of nonreciprocal coupling, we will fix the onsite energy as $w_A=5~$GHz, $w_B = 6~$GHz, and $w_C = 7~$GHz respectively, until otherwise claimed explicitly.

\section{Symmetry and topology}
With the time reversal operation represented by complex conjugation $K$, the NH Hamiltonian, Eq.~(\ref{eqht}) is $\mathcal{PT}$ symmetric, where the parity operator is explicitly given as

\begin{gather}
\label{P}
\mathcal{P} =
\begin{pmatrix}
1 & 0 & 0 \\
0 & 1 & 0 \\
0 & 0 & -1
\end{pmatrix}.
\end{gather}

As a result of $\mathcal{P}$ being unitary, $\mathcal{PT}$ symmetry dictates that any non-real eigenvalue is accompanied by its complex-conjugate pair. Consequently, for any odd dimensional $\mathcal{PT}$ Hamiltonian there is always a pure real eigenvalue as shown in Fig.~\ref{figic}. 
Thus, $\mathcal{PT}$ symmetry imposes an equivalent balanced gain–loss structure on the trimer, constraining the eigenvalues to evolve as real values or complex-conjugate pairs. 

A third-order exceptional point ($\text{EP}_3$) is realized when all three eigenstates of the Hamiltonian coalesce simultaneously with their corresponding eigenvalues. Such an $\text{EP}_3$ can be achieved by coalescing two second-order exceptional points ($\text{EP}_2$). 
Deliberate numerical analysis indicates that for $\Omega_1 < \Omega_p = \frac{1}{\sqrt{2}} \text{ GHz}$,  three distinct $\text{EP}_2$—labeled $a_1, a_2,$ and $a_3$—formed by pairwise coalescence (Figs. \ref{6d} and \ref{6a}) can be obtained by varying $\Omega_2$. In particular, the close proximity and distinction between $a_1$ and $a_2$ makes this possible for investigating inter-EP interactions, as will be discussed later.
These three EPs divide the energy bands into four regions, each of which corresponds to a unique dynamical phase of the system.
The symmetric phase is realized with parameters satisfying $a_2 > \Omega_2 > a_1$ and $\Omega_2 > a_3$. 
Here, the energy gaps are purely real, yielding Rabi-like dynamical oscillations.
Conversely, for $\Omega_2 < a_1$ or $a_3 > \Omega_2 > a_2$, the imaginary gap emerges and the system exhibits spontaneous symmetry-breaking characteristics, where the system evolves towards a phase-locked steady state. 

As the imaginary coupling strength $\Omega_1$ increases, the gap between $\text{EP}_2$ $a_1$ and $a_2$ diminishes gradually. As $\Omega_1$ approaches the critical value $\Omega_p$, $\text{EP}_2$ $a_1$ and $a_2$ coalesce to a single point $a_1 = a_2 = a_p$. This is exactly the $\text{EP}_3$ we promised earlier, which is characterized by a $k^{1/3}$-type dispersion relation~\cite{mandal2021symmetry}. In this situation, the symmetric phase located in the interval $(a_1,a_2)$ between the two symmetry-broken phases is squeezed out, as illustrated in Figs. \ref{6e} and \ref{6b}). 
\begin{figure}[t]
\centering
\includegraphics[width=0.4\textwidth]{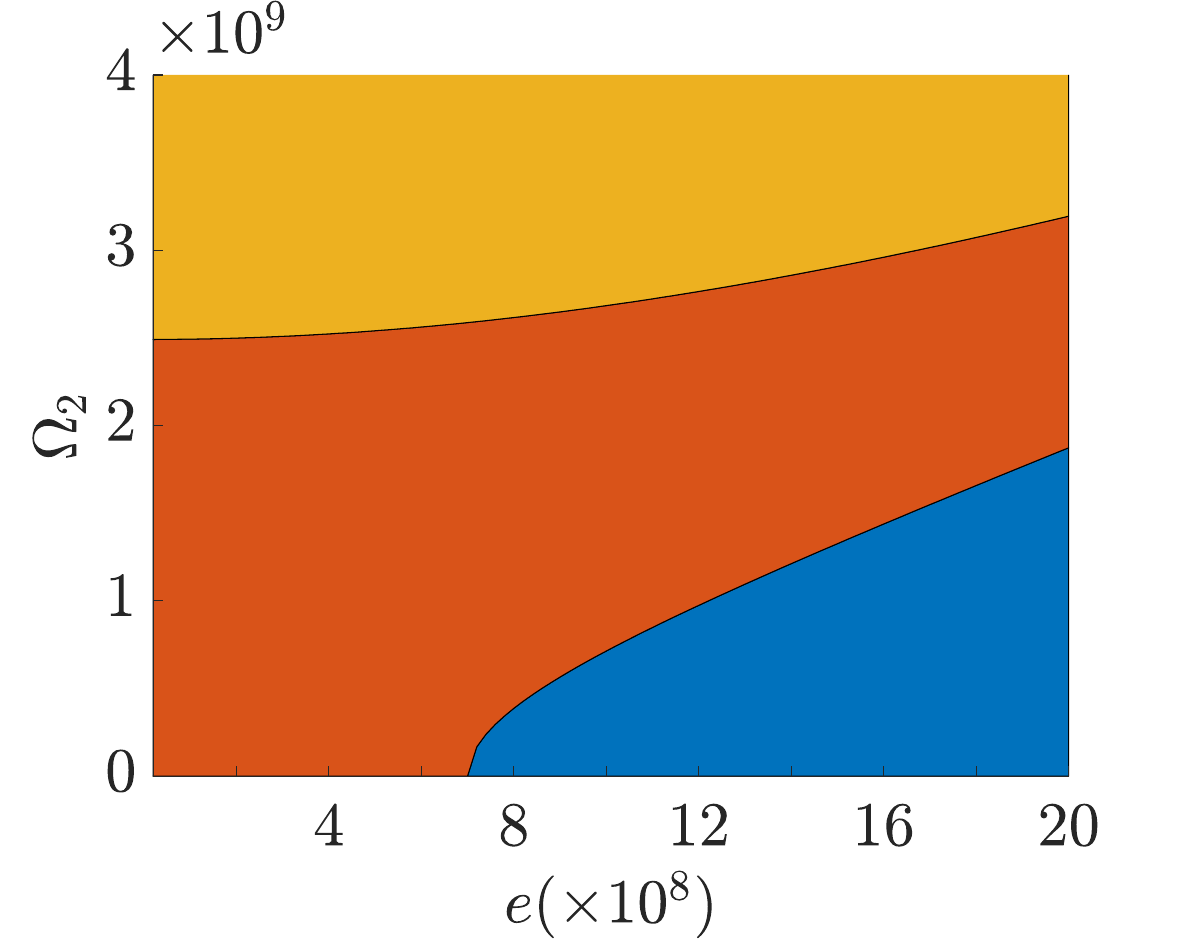}
\caption{\label{phdia}Phase diagram of $\text{EP}_2$ and $\text{EP}_3$. The upper and lower boundary curves correspond to second-order and third-order EPs, respectively. The dynamical evolution in yellow region corresponds to symmetry-unbroken phase, while the other two regions represent distinct symmetry-broken phases connected by a third-order exceptional phase transition.}
\end{figure}

The landscape of the dynamical phase is demonstrated as in Fig.~\ref{phdia} for parameter space $e-\Omega_2$. The whole diagram is partitioned into three distinct dynamical phases by two boundary curves, which represent $\text{EP}_2$ and third-order exceptional points ($\text{EP}_3$), respectively. Specifically, $\text{EP}_2$ mediate the phase transition between the symmetry-unbroken phase (yellow region) and one symmetry-broken phase (red region), while $\text{EP}_3$ govern the phase transition between the two distinct symmetry-broken phases (red and blue regions). Numerical analysis shows that to observe the appearance of $\text{EP}_3$, the imaginary coupling strength has been fixed at $\Omega_1 = \Omega_p$. As demonstrated, a ${\rm EP}_2$ is almost always approachable. However, the appearance of ${\rm EP}_3$ requires a relatively high degree of nonreciprocal $e$. It can be shown that the critical imaginary coupling strength $\Omega_p$ depends solely on the onsite energies, which can be given analysis as $\Omega_p = \pm\sqrt{\frac{\delta_C^3}{\delta_C-\delta_A}}$ where $\delta_i=w_i-\overline{w}$ ($i=A,B,C$), and $\overline{w} = \frac{w_A+w_B+w_c}{3}$. When $\Omega_1 = \Omega_p$, the line of $EP_3$ can be obtained analytically by the following equation
\begin{align}
\Omega_2^2-e^2 = \frac{\delta_A^3}{\delta_C-\delta_A}.
\end{align}


For imaginary coupling $\Omega_1 > \Omega_p$, the two symmetry-broken phases merge into a single phase with the intervening third-order exceptional phase transition eliminated (Fig. \ref{6f}, \ref{6c}). 
As we will demonstrate later, although the 3rd-order EP disappears, operating in the near-EP parameter regime leaves a residual effect on the entanglement dynamics. 
In the limit $\Omega_1 \rightarrow \infty$, the band structure of the system becomes similar to a two-level system and can only have second order degeneracy.

\section{Signature of quantum phase transitions:  fidelity}
Quantum phase transitions give rise to quantum criticality, which in turn manifests itself in a variety of phenomena. These relevant studies greatly advance our understanding of the mechanism of organization in the quantum world.
To illustrate the plausible qualitative difference among the regions around the EPs, in this section, we provide a numerical analysis based on the concept of fidelity~\cite{zanardi2006ground,tang2021unveiling}. Unlike entanglement, which mainly dictates the nonlocal correlation between different partitions, fidelity tells how the state of the systems changes according to the adjustment of the relevant parameters. Intuitively, it is more natural to delineate whether a change is gradual or critical. Indeed, fidelity anomalies can correlate with peaks in the generation of entanglement, demonstrating how quantum criticality dictates entanglement dynamics across the parameter space \cite{li2023speeding,yuan2026beating}.

\begin{figure*}[t]
    \begin{center}
        \subfigure[\;$\Omega_1=6\times10^8$]{\includegraphics[width=0.23\textwidth]{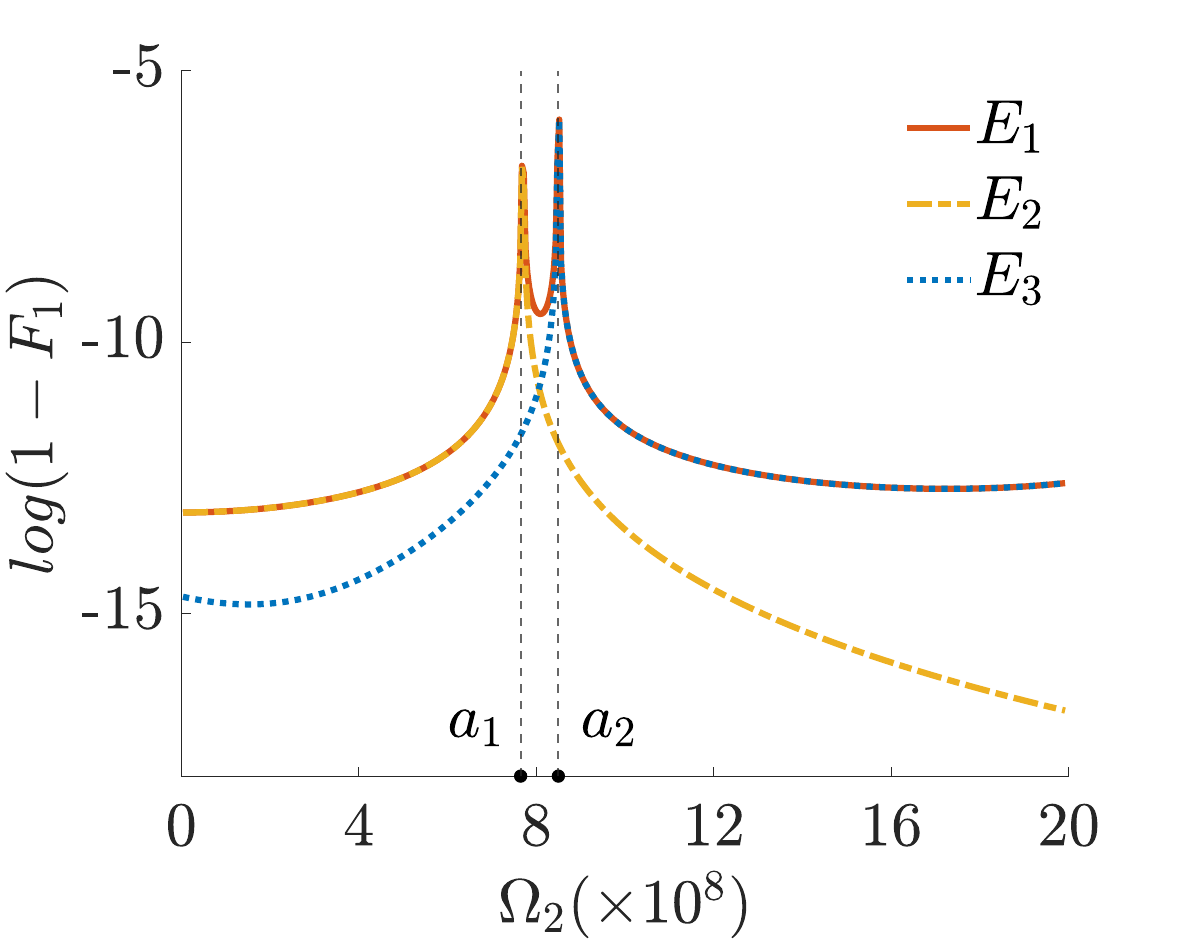}
        \label{f16}}
        \subfigure[\;$\Omega_1=\Omega_p$]{\includegraphics[width=0.23\textwidth]{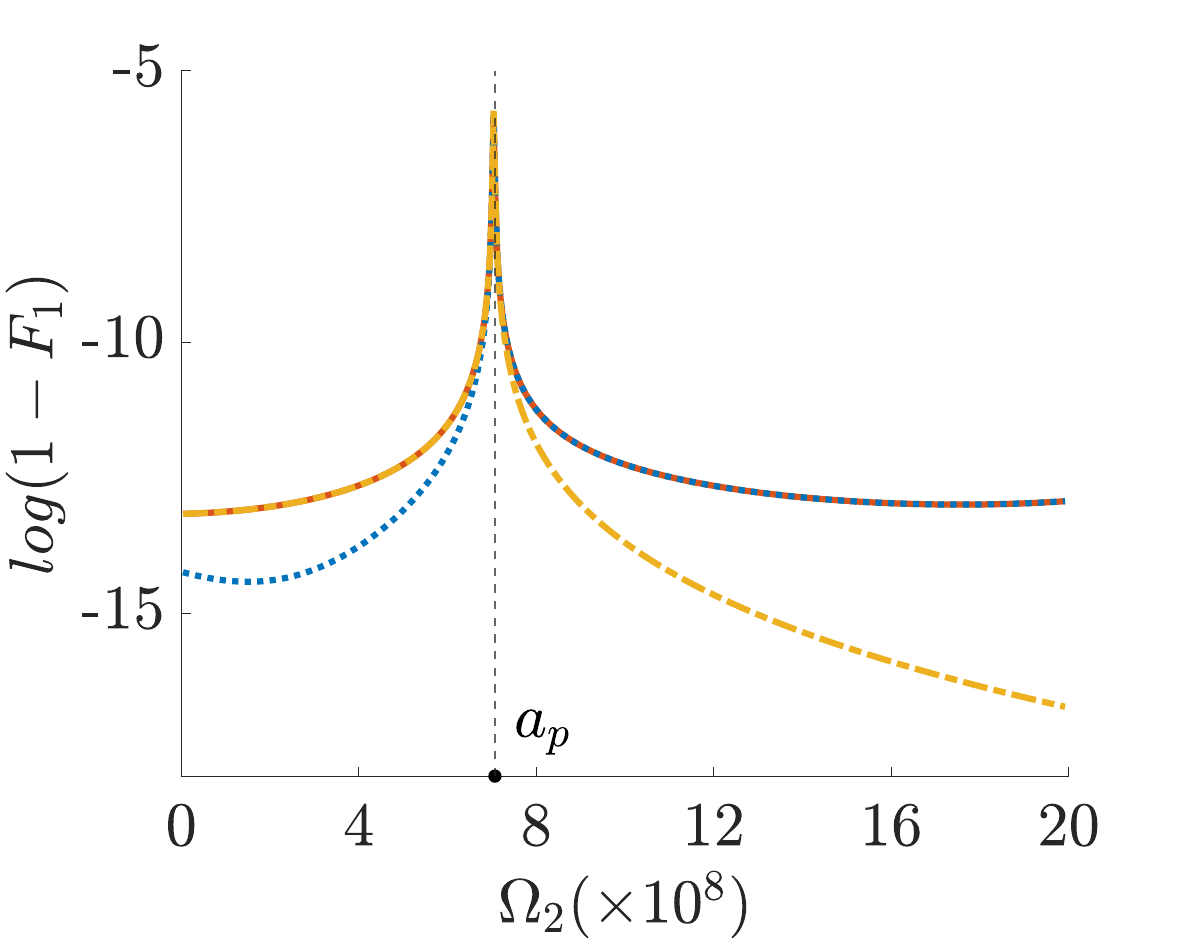}
        \label{f1p}}
        \subfigure[\;$\Omega_1=6\times10^8$]{\includegraphics[width=0.23\textwidth]{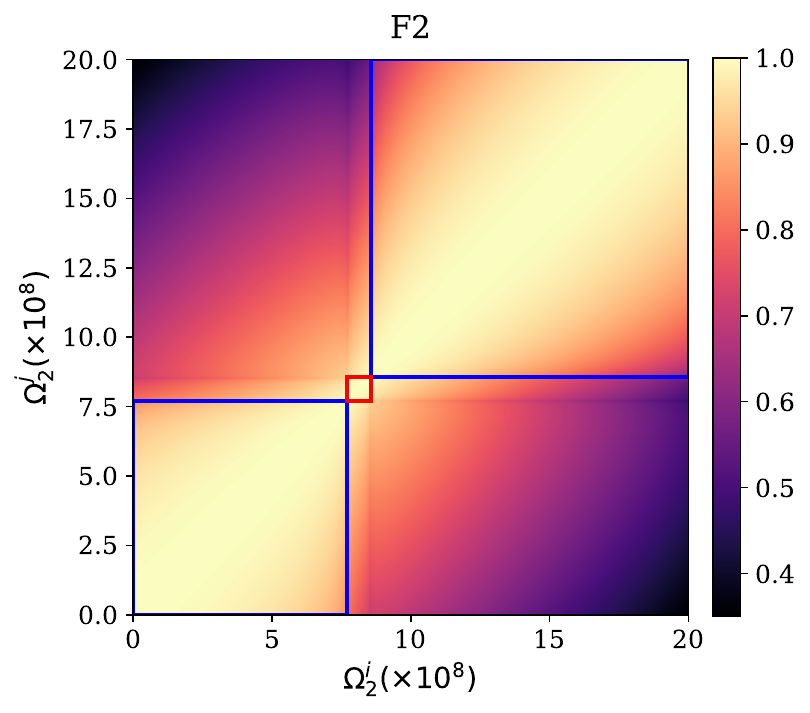}
        \label{f26}}
        \subfigure[\;$\Omega_1=\Omega_p$]{\includegraphics[width=0.23\textwidth]{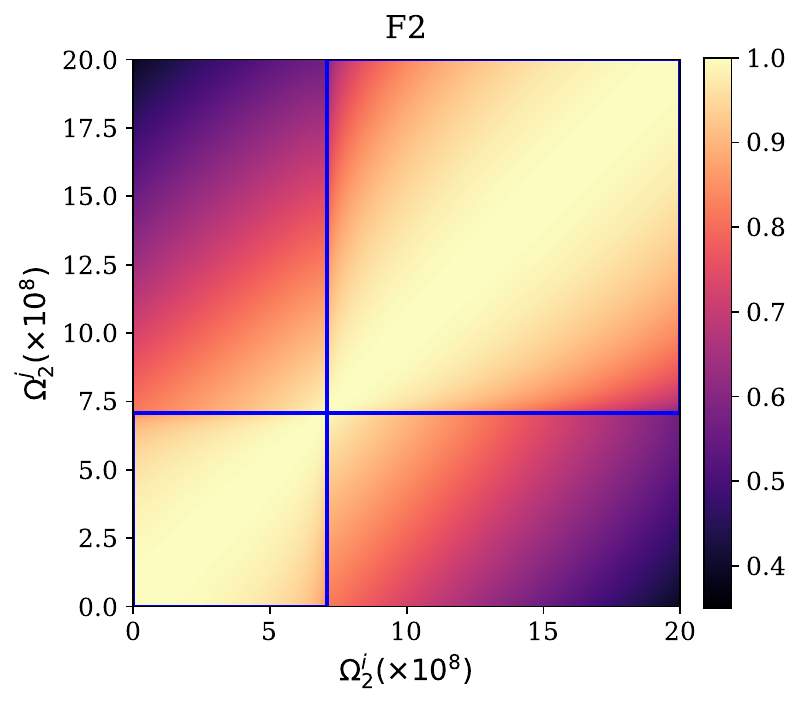}
        \label{f2p}}
        \caption{\label{fig:fidelity}Fidelities of the eigenstates in the NH three-level system. With varying coupling coefficient $\Omega_2$, we construct the fidelity in two kinds of definitions. (a)-(b) First kind of fidelity with $\delta\Omega_2=10^6 Hz$. Red, yellow and blue curves represent fidelities of the three eigenstates fidelity. (c)-(d) Second kind of fidelity. Light yellow indicates high fidelity region while the dark purple indicates low fidelity. The squares with blue outlines are the calculated symmetry-broken phase region and the square with red outlines is the symmetry-unbroken phase.}
    \end{center}
\end{figure*}

The conventional fidelity is defined as the overlap between states of neighboring driving parameter $F_1=|\braket{\phi(\lambda)|\phi(\lambda+\delta\lambda)}|$, here $\ket{\phi}$ denote the eigenstates of the model and $\lambda$ represents the driving parameter (taken as $\Omega_2$ in this work). The fidelity $F_1$ for each eigenstate is plotted in Figs. \ref{f16} and \ref{f1p} with a small scaling of $\delta\Omega_2=1~$MHz. Compared with non-Hermitian energy level spectra (Fig. \ref{figic}), this kind of fidelity exhibits excellent agreement with the spectral features of the system, clearly identifying the positions of phase transition points with the peaks in function $log(1-F_1)$. Specifically, $E_1$ (red curve) reveals all phase transition points, while $E_2$ (yellow curve) and $E_3$ (blue curve) exhibit a single peak at $a_1$ and $a_2$, respectively. At critical coupling $\Omega_1 = \Omega_p$, three eigenstates undergo phase transitions simultaneously, the peaks at $a_1$ and $a_2$ in all three eigenstate fidelities merge into a single peak at $a_p$,which corresponds to an $\text{EP}_3$. 

To enhance the visibility of phase boundaries and extract richer structural information from the NH system, we employ an alternative fidelity measure referred to as the fidelity map, given by $F_2=|\braket{\phi(\lambda_i)|\phi(\lambda_j)}|$. This quantity quantifies the similarity between eigenstates at two different values of the driving parameter $\lambda_i$ and $\lambda_j$,  providing a two-dimensional representation of the phase transitions. We plot the fidelity map $F_2$ for the eigenstate $E_1$ in Figs. \ref{f26} and \ref{f2p}, as it is sufficient to analyze all phase transitions. The two-dimensional color maps exhibit distinct square regions that provide a clear identification of different phases and transition points. Different square regions exhibit different characteristic patterns and behaviors corresponding to the symmetric phase and the symmetry-broken phase, respectively.

Although fidelity offers a straightforward and visually clear characterization, its limitations are also evident. In an NH three-level system, the gain eigenstates may evolve into a decay eigenstate as the driving parameter varies. As a result, the tracked states can decay rapidly during certain stages of the phase transition, making them difficult to measure and thus not robust. Furthermore, the fidelity-based approach lacks an intrinsic physical quantity that can unambiguously distinguish higher-order EPs from $\text{EP}_2$. In practice, the presence of high-order EPs can only be inferred from the merging of multiple phase-transition points, which inevitably limits the accuracy of the method. To address these issues, we introduce an entanglement-based method that enables the study of the intrinsic quantum-mechanical properties of the system and its higher order characteristics.

\section{Entanglement Phenomena at High-Order Exceptional Points}
\begin{figure}[t]
    \begin{center}
        \subfigure[\;$\Omega_2=7\times10^8$]{\includegraphics[width=0.23\textwidth]{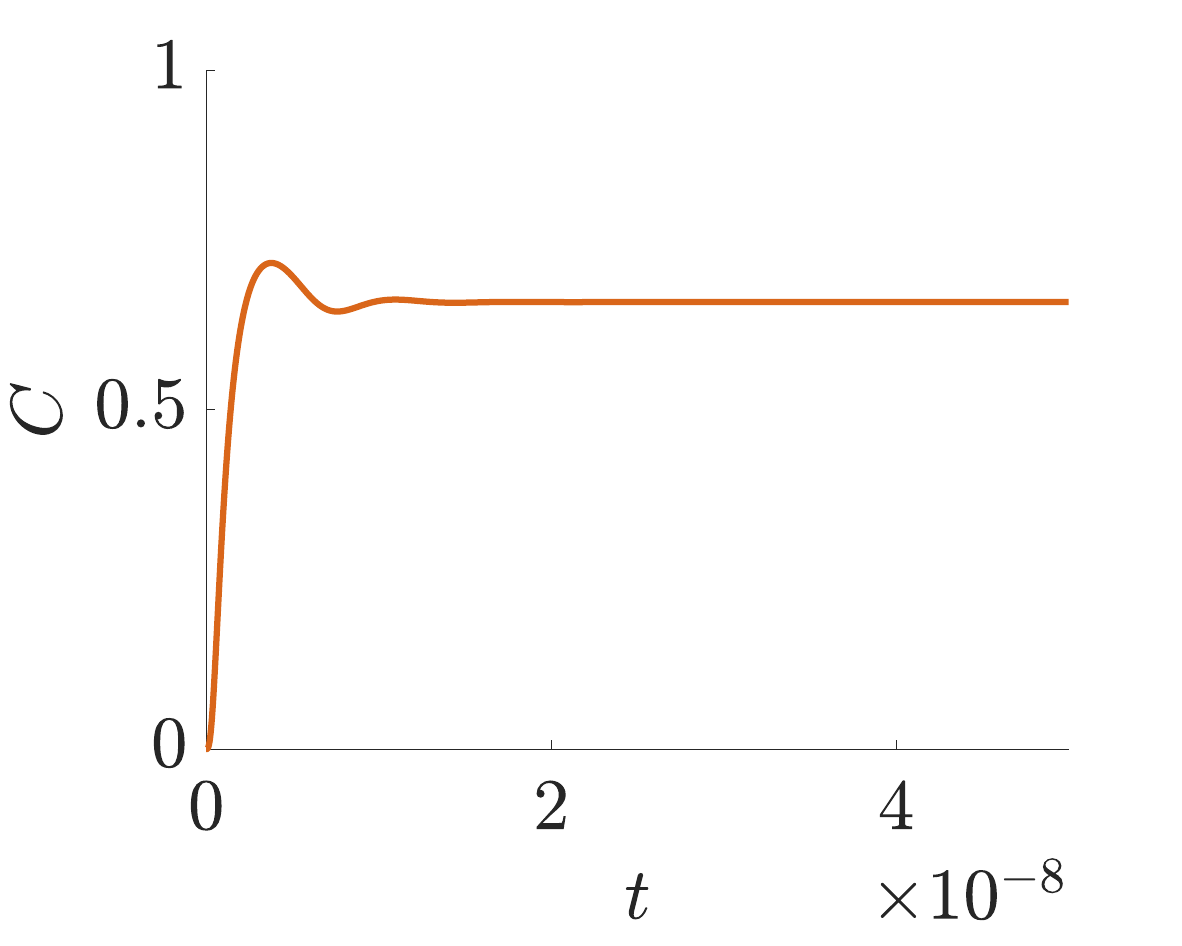}
        \label{con7}}
        \subfigure[\;$\Omega_2=8\times10^8$]{\includegraphics[width=0.23\textwidth]{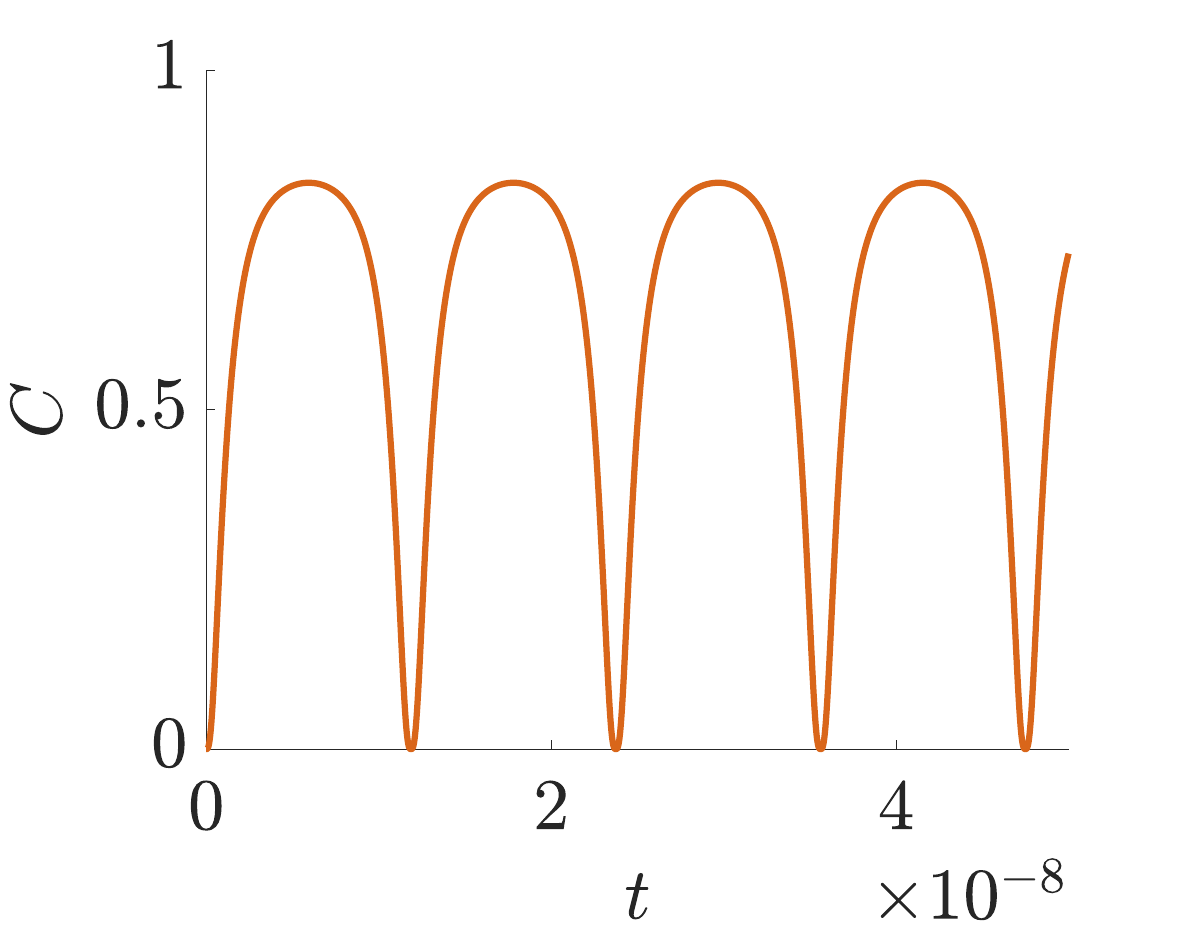}
        \label{conr8}}
        \subfigure[\;$\Omega_2=9\times10^8$]{\includegraphics[width=0.23\textwidth]{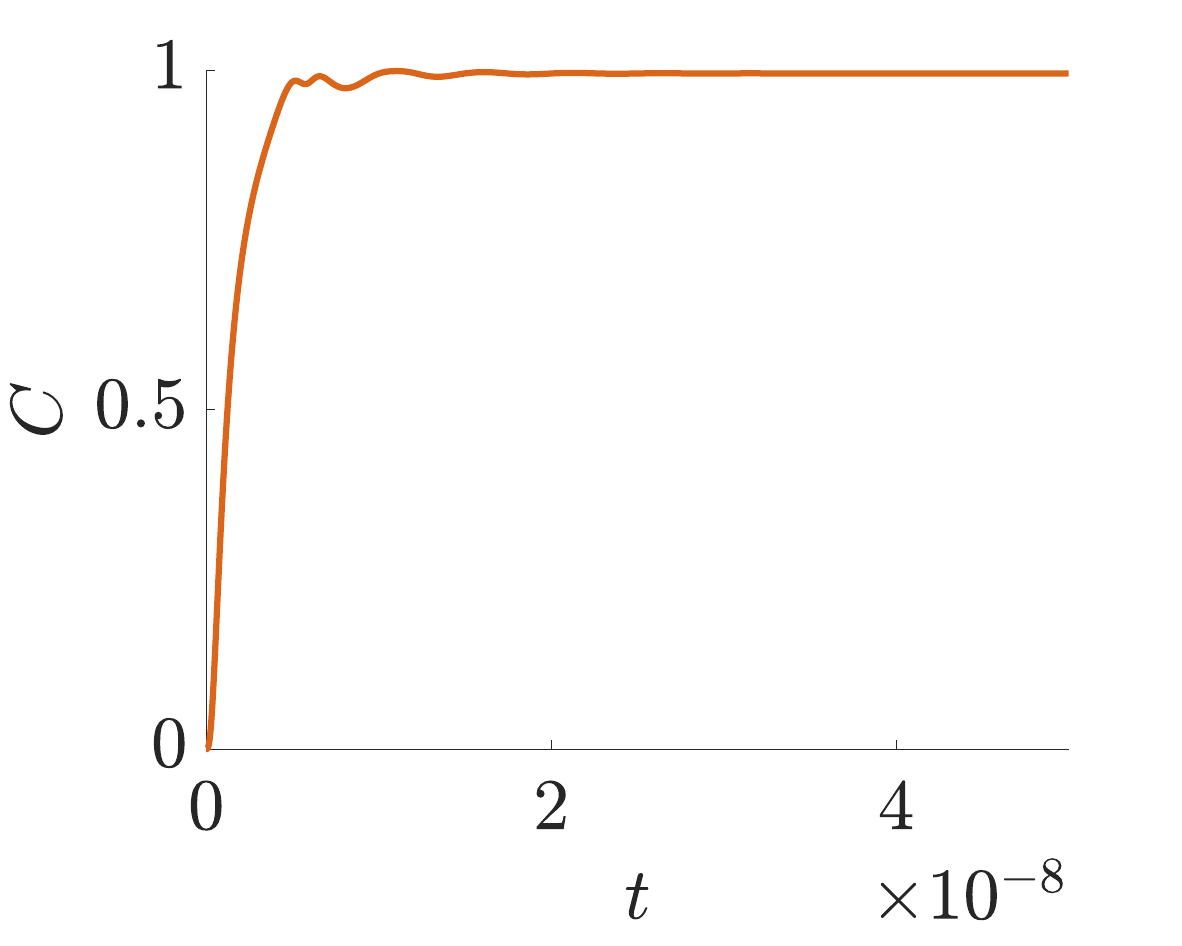}
        \label{con9}}
        \subfigure[\;$\Omega_2=30\times10^8$]{\includegraphics[width=0.23\textwidth]{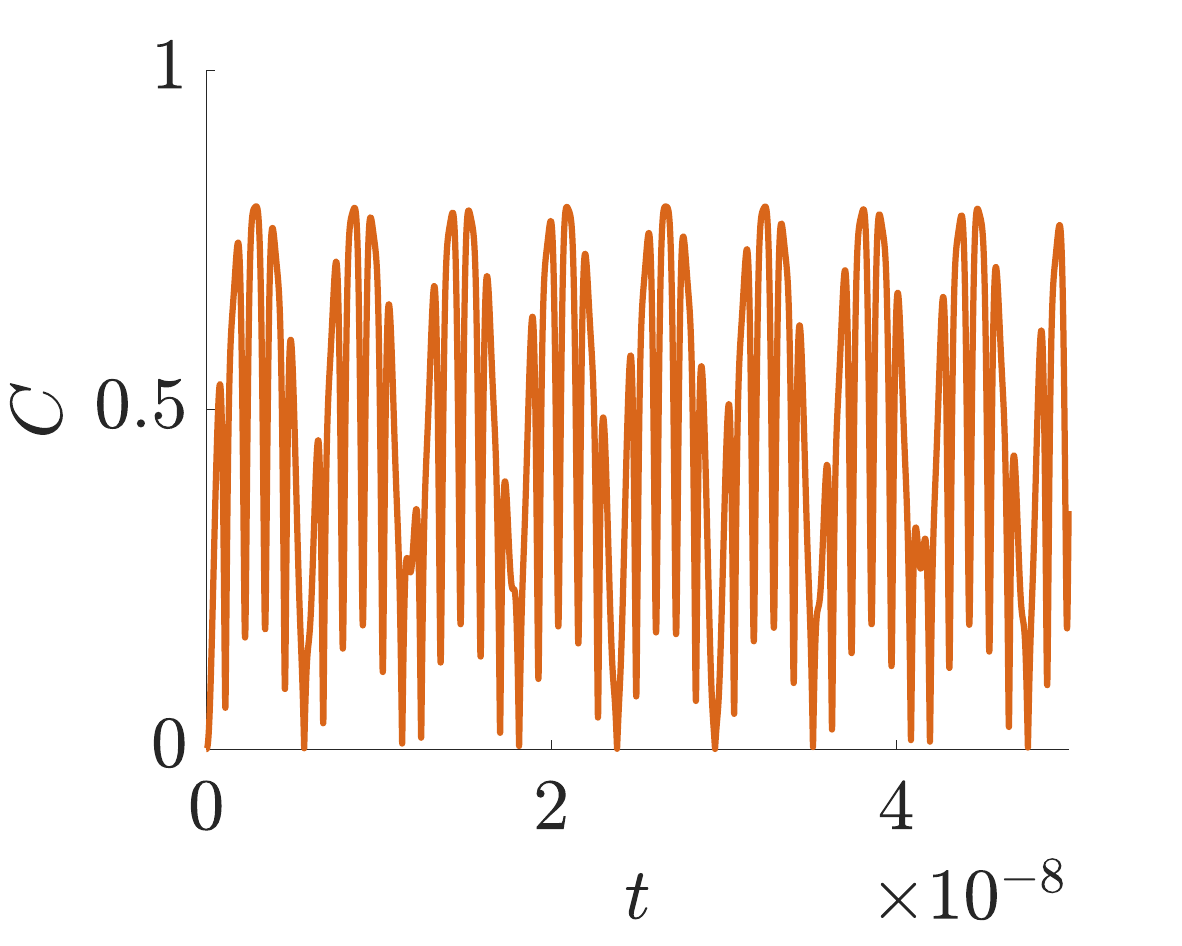}
        \label{con30}}
        \caption{\label{fig:cone}Temporal evolution of concurrence for $\Omega_1 < \Omega_p$. (a),(c) The concurrence relaxes to a constant value $C_{ss}$ at late times, corresponding to the dynamical symmetry-broken phase. (b),(d) The concurrence exhibits persistent Rabi-like oscillatory dynamics over time, corresponding to the dynamical symmetry-unbroken phase.}
    \end{center}
\end{figure}

To support the discovery above, in this section we present a detailed study of the entanglement dynamics. It is well known that for two interacting quantum bodies, as time passes, they will get entangled with each other. Entanglement is not only an important sort of quantum resource, it is also a particular useful key to the fundamental problem in quantum many-body systems. To quantitatively account for the amount of entanglement between two cavities \cite{wootters1998entanglement}, we use concurrence, which can be obtained as
\begin{gather}
\label{eqc}
C_{ij}=\frac{2|A_i||A_j|}{|A|^2}.
\end{gather}
Where the subscripts $i$ and $j$ label two of the cavities, where $A_i$ and $A_j$ are the amplitudes of the states $|i\rangle$ and $|j\rangle$, respectively. $|A|^2$ denotes the sum of the squares of the amplitudes of all three states $\ket A, \ket B, \ket C$ (assuming pure state), which gives the normalization coefficient. We will only consider the concurrence $C_{BC}$ between cavity B and cavity C which is simplified as $C$. See Appendix \ref{appendix a} for details about the derivation. For $\Omega_1 < \Omega_p$, the system presents only $\text{EP}_2$. Thus, we can employ the same method as the dissipative JC model \cite{han2023exceptional} to characterize the dynamical phase transition. The dynamical phases can be distinguished by the evolution of concurrence, since its behavior closely follows the temporal evolutions of the amplitudes of the states. As in the region where $a_1 < \Omega_2 < a_2$ (e.g. $8 \times 10^8$ in Fig. \ref{6d}) and in the region where $\Omega_2 > a_3$, the concurrence exhibits oscillatory dynamics similar to Rabi oscillations, as illustrated in Figs. \ref{conr8} and \ref{con30}. In other regions, it will evolve to a fixed value $C_{ss}$ in steady state, as shown in Figs. \ref{con7} and \ref{con9}.
When $\Omega_1 = \Omega_p$, the interval $(a_1,a_2)$ shrinks to a single point, which is a third-order degenerate EP. This third-order degenerate EP can potentially serve as a phase transition point connecting two distinct symmetry-breaking phases. Due to the nature of this kind of dynamical phase, we focus on studying the steady state entanglement phenomenon to determine the characteristics of the third-order degenerate EP phase transition.

\begin{figure*}[t]
    \begin{center}
        \subfigure[\;]{\includegraphics[width=0.3\textwidth]{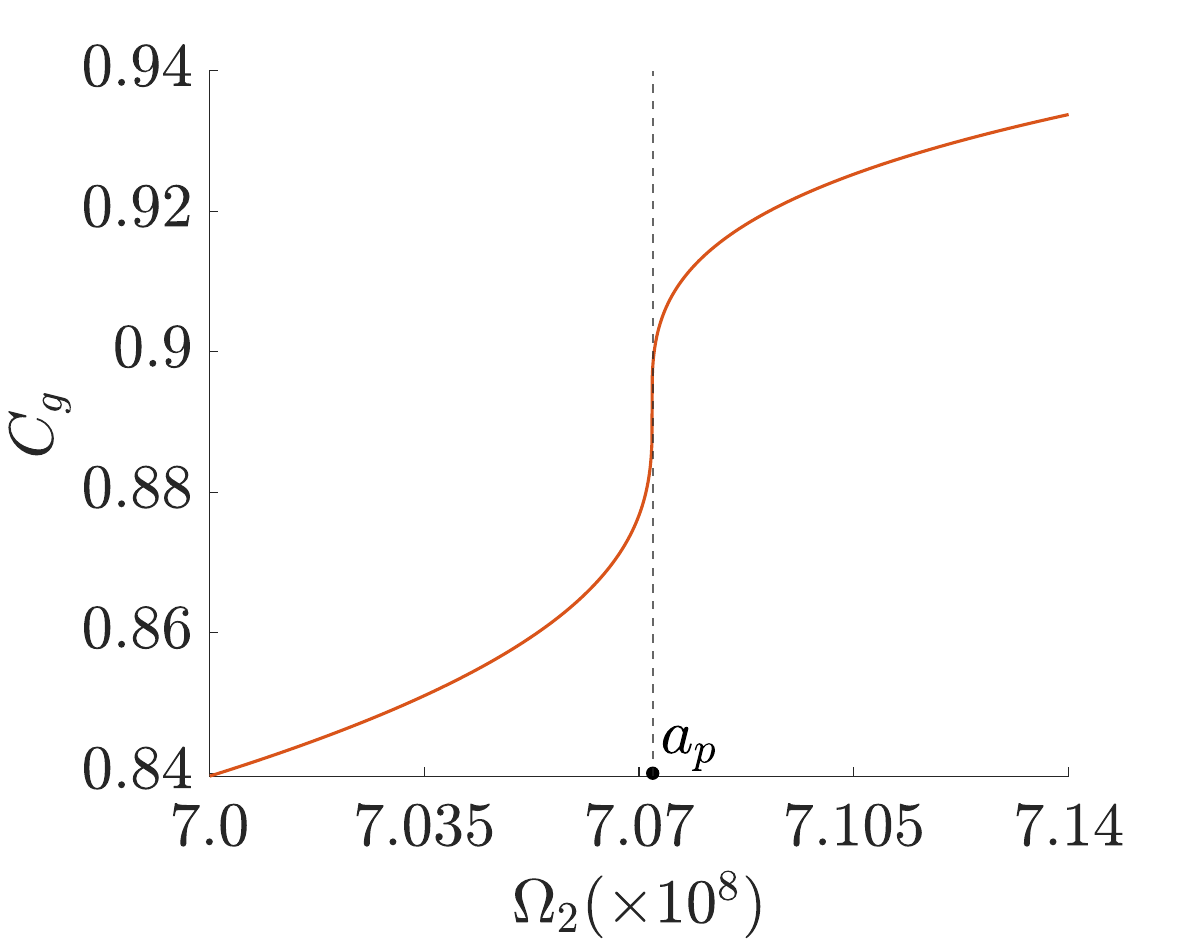}
        \label{h1}}
        \subfigure[\;]{\includegraphics[width=0.3\textwidth]{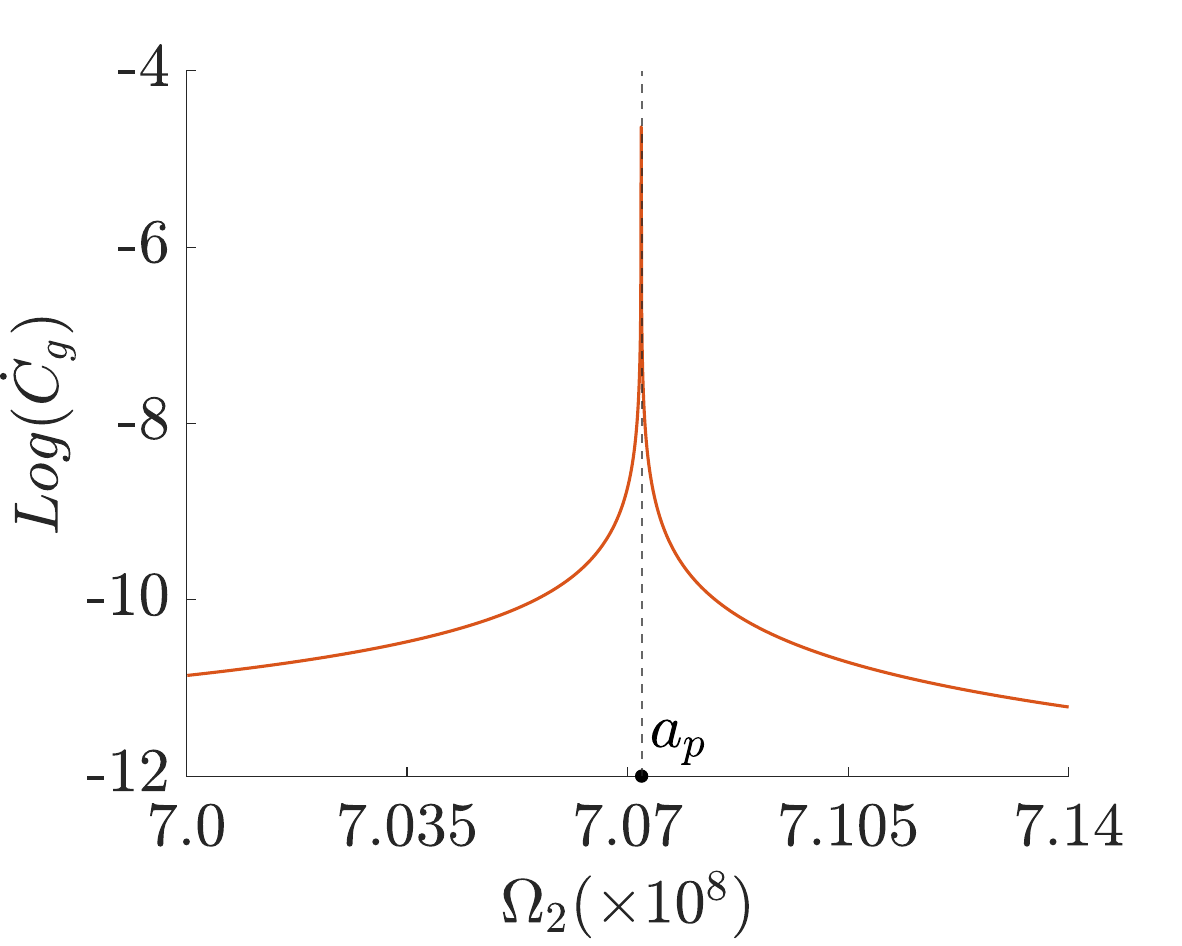}
        \label{hcon}}
        \subfigure[\;]{\includegraphics[width=0.3\textwidth]{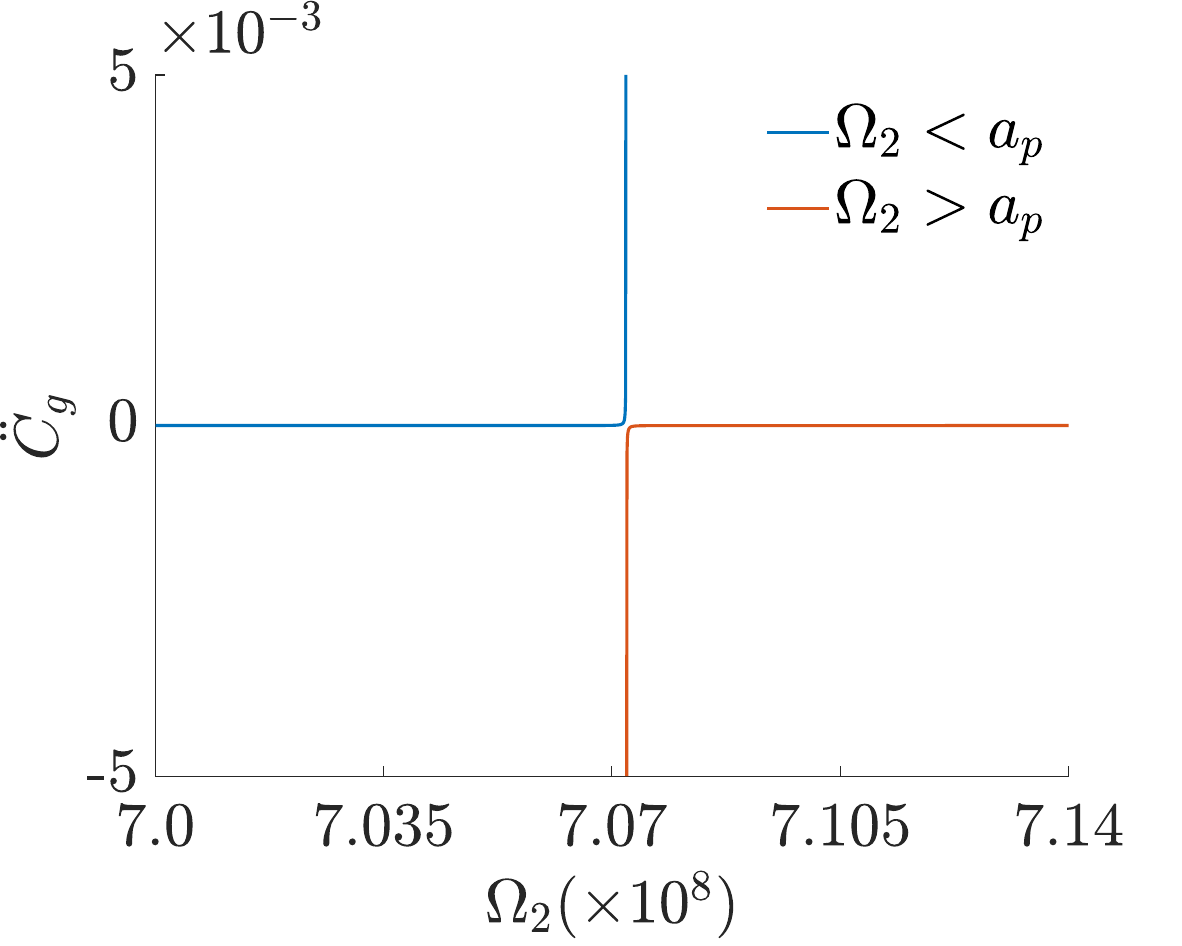}
        \label{hk3}}
        \caption{\label{fig:conss}Entanglement-based characterization of third-order EP phase transition point ($\Omega_1=\Omega_p$). (a)Gain-state concurrence $C_g$ as a function of$\Omega_2$. It is qualitatively different with second-order EP phase transition. (b)Logarithm of the absolute first derivative of the concurrence $Log(\dot C_g)$ versus $\Omega_2$. When phase transition occurs, the first derivative exhibits a steep peak near the EP, and it is continuous at the peak. (c) $\ddot C_g-\Omega_2$. It has a non-zero change at the phase transition point.}
    \end{center}
\end{figure*}

To investigate the relation between entanglement and EP dynamical phases, we employ the steady-state concurrence, or equivalently the gain mode concurrence $C_g$ (see Appendix \ref{appendix b}). The gain modes for different $\Omega_2$ are indicated in Fig. \ref{figic}. As demonstrated in Fig. \ref{figic}, the eigenstate originally associated with the gain modes is replaced by another eigenstate as $\Omega_2$ varies. This is the primary motivation for us to use the steady-state concurrence instead of eigenstate concurrences \cite{han2023exceptional}. We focus on $\text{EP}_3$ which arises specifically when $\Omega_1 = \Omega_p$. In this case, the concurrence of the gain mode $C_g$ exhibits a sharp increase near the $\text{EP}_3$ (Fig. \ref{h1}). However, in contrast to the case of second-order degeneracy, the first derivative of $C_g$ with respect to $\Omega_2$ does not show pronounced discontinuity at the phase transition point, which can be an indicator to distinguish $\text{EP}_2$ and  $\text{EP}_3$. Nevertheless, the derivative of concurrence exhibits an intense soar and dive with continuous change, as shown in Fig. \ref{hcon}.  This suggests that this $\text{EP}_3$ is potentially a continuous phase transition point. To verify this, we calculate the second derivative of the concurrence $d^2C_g/d\Omega_2^2$, which is plotted in Fig. \ref{hk3}. The red and blue curves correspond to the regions where $\Omega_2 < a_p$ and $\Omega_2 > a_p$, respectively. A noticeable jump appears in the second derivative curve at the phase transition point $a_p$. This observation further implies that the third-order EP corresponds to a continuous phase transition point. Additionally, near the phase transition point, $C_g$ exhibits a sensitive response to variations in $\Omega_2$, which is potential for future applications such as quantum sensing.

When $\Omega_1=\Omega_p$, an $\text{EP}_3$ emerges, which represents a continuous phase transition point. In this regime, the system exhibits two phase transition points with respect to $\Omega_2$: one at $a_p$, corresponding to a second-order dynamical phase transition, and the other at $a_3$, associated with a first-order dynamical phase transition. Consequently, three distinct dynamical phases emerge in this scenario, corresponding to the phase diagram in Fig. \ref{phdia}.
Thus, we can further recognize the structure of the phase diagram. The phase below the continuous phase transition curve corresponds to a symmetry-broken phase dominated by asymmetric coupling. The intermediate phase corresponds to a symmetry-broken phase dominated by imaginary coupling. The phase above the curve of $\text{EP}_2$ is the symmetry-unbroken phase. $\text{EP}_3$ vanish when $e$ is sufficiently small. This occurs for the reason that the NH effect induced by $e$ in asymmetric coupling is much weaker than the imaginary coupling. Consequently, the leading factor for symmetry-breaking in the system remains to be the imaginary coupling, and there is no continuous phase transition as $\Omega_2$ varies anymore.

\section{EP-enhanced entanglement dynamics in coupled third order EP Non-hermitian system}
\begin{figure*}[t]
    \begin{center}
        \subfigure[\;$\Omega_1=3\times10^8$]{\includegraphics[width=0.23\textwidth]{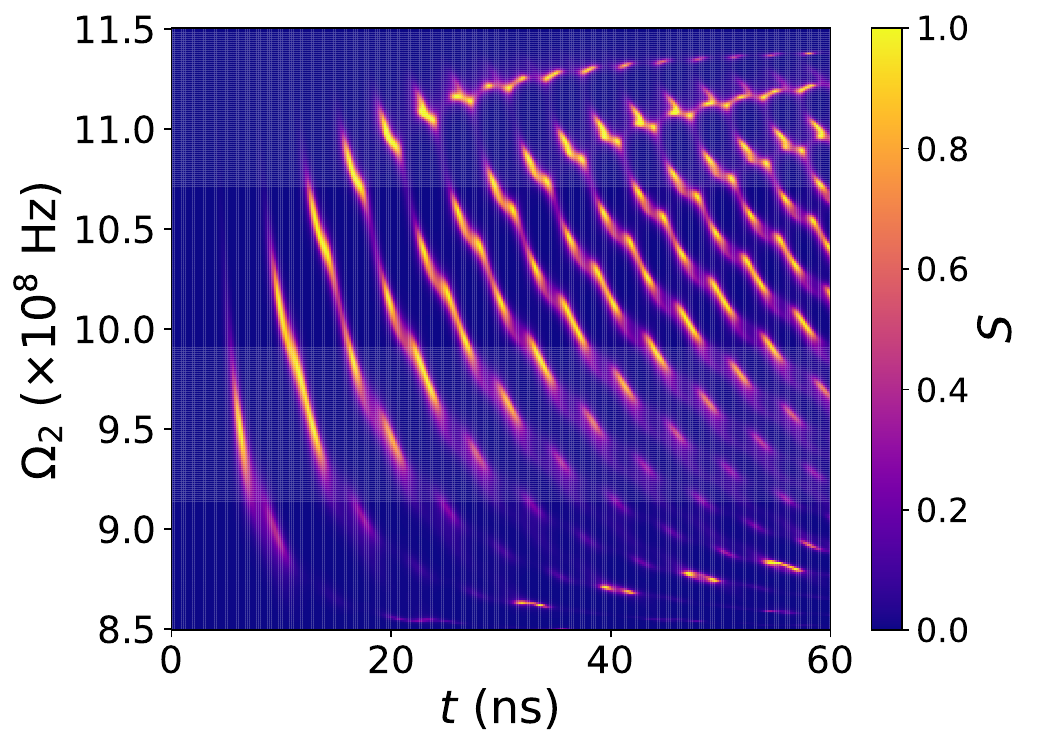}
        \label{osp}}
        \subfigure[\;$\Omega_1=\Omega_p$]{\includegraphics[width=0.23\textwidth]{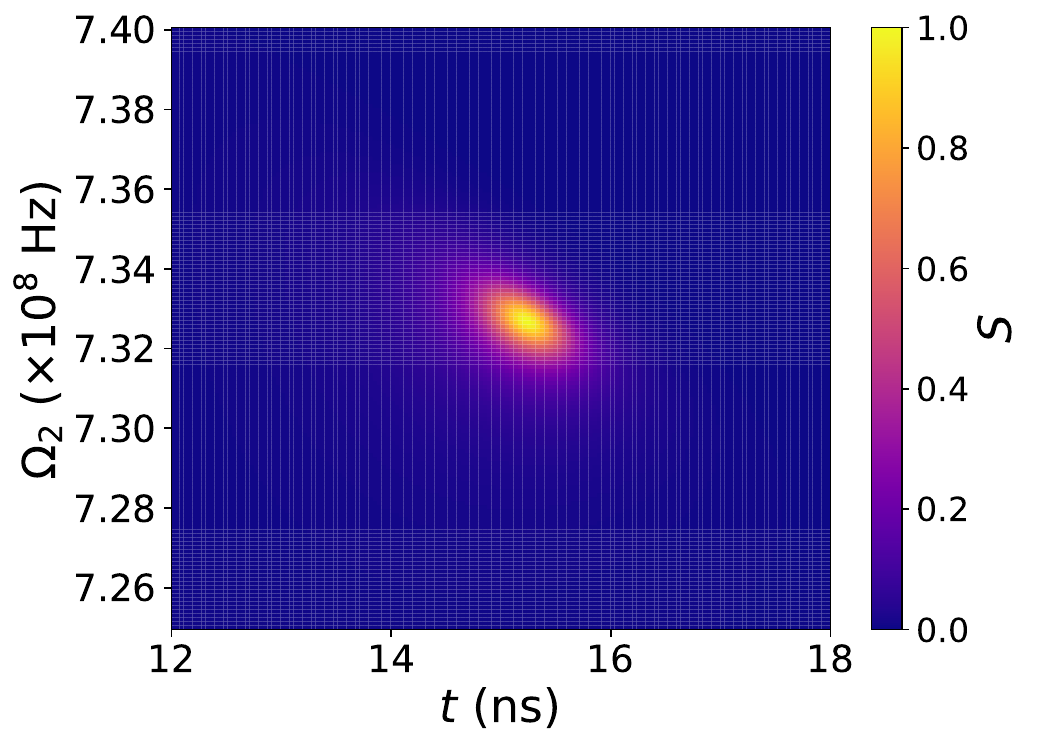}
        \label{oep}}
        \subfigure[\;$\Omega_1=7.2\times10^8$]{\includegraphics[width=0.23\textwidth]{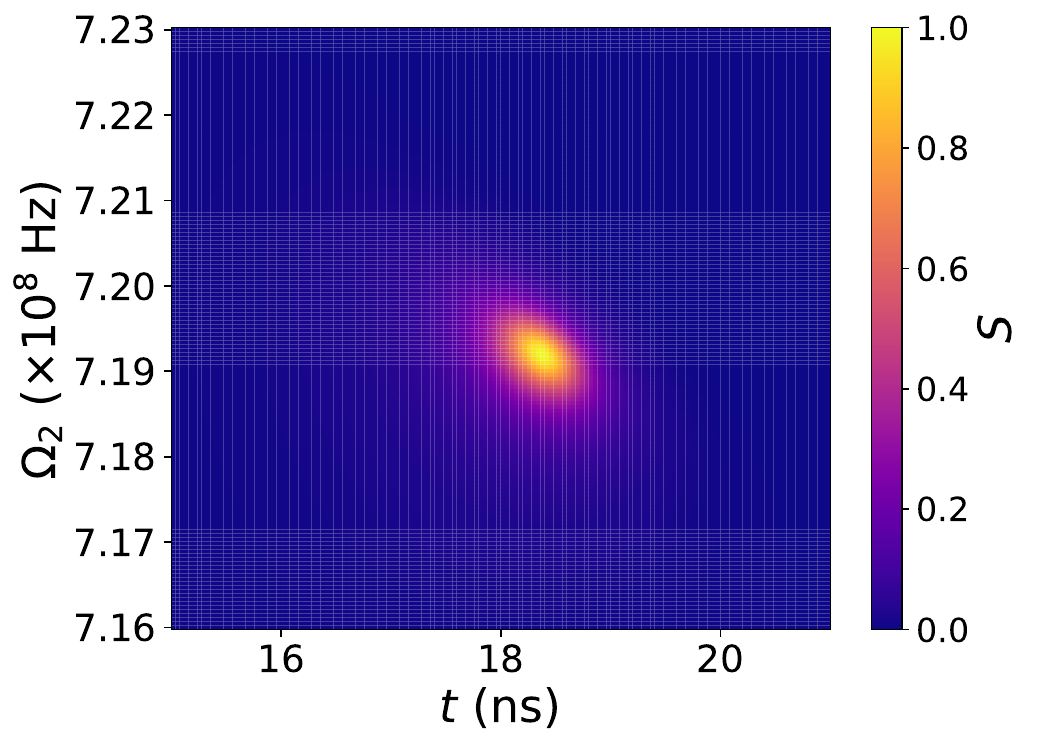}
        \label{olp}}
        \subfigure[]{\includegraphics[width=0.23\textwidth]{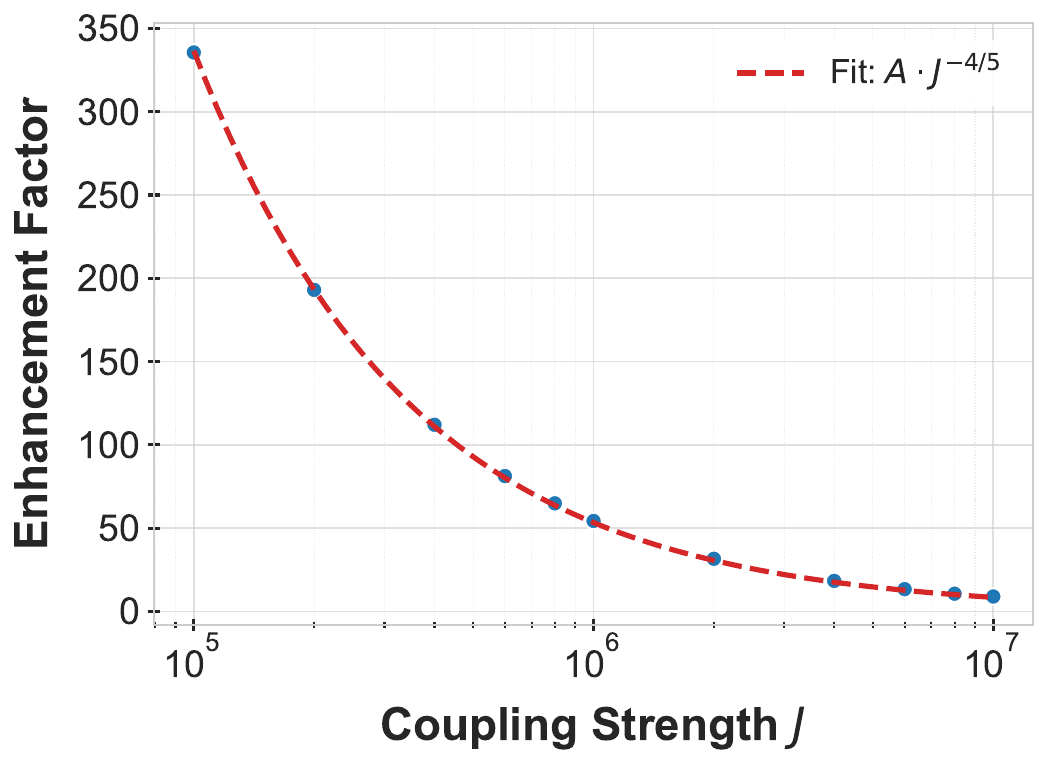}
        \label{enhance}}
        \caption{\label{fig:fidelity}(a)-(c) Entanglement entropy evolution as a function of $\Omega_2$ in three different topological phases. Here we set $w_{A}=50\times10^8, w_{B}=60\times10^8, w_{C}=70\times10^8, e=10\times10^8$ and $J=0.1\times10^8$ (Hz). (d) Entanglement enhancement factor defined as ratio of first time to reach first maximum entropy for Hermitian system ($J^{-1}$) to the corresponding time $t_m$ for our NH system.}
    \end{center}
\end{figure*}
It was demonstrated that the speed of entanglement generation can be accelerated even without limitation by exploiting the NH property compared to the Hermitian case~\cite{li2023speeding, yuan2026beating}; however, this happens only with the availability of an unlimited tunneling amplitude.  Recent work with two identical two-level systems show that a 4th order EP can facilitate the generation of entanglement: by weakly coupling ($J$) the two two-level systems, an energy gap is opened at the 4th order EP structure, leading to an acceleration of $J^{1/4}$ compared to the standard Hermitian $J$. This raises a series of interesting questions: is the acceleration generally true for any higher order EPs? What about the case for nonreciprocally coupled systems for which the mathematics is quite different, and even its Hermitian counterpart can not be defined easily? Are there some new especially with the presence of multiple EPs? To explore this, we construct the three-cavity setup introduced in the previous chapter as effective three level system and couple two identical subsystems through two of their energy levels

\begin{align}
\label{couple}
H=\sum_{j=1,2}H_j+J(\sigma_+^{(1)}\sigma_-^{(2)}+\sigma_-^{(1)}\sigma_+^{(2)}),
\end{align}
where $H_j$ refer to two same subsystems with the same Hamiltonian in Eq. \ref{eqht}, $\sigma_+=a_B^\dagger a_A$ and $\sigma_-=a_A^\dagger a_B$. Here, 1 and 2 refer to the two subsystems and $J\ll E_n$. To quantify the entanglement generated between these two subsystems, we use the entanglement entropy $S$, defined as

\begin{align}
\label{entropy}
&S(t)=-\operatorname{Tr}[\rho_1(t)\log_2\rho_1(t)],
\end{align}
where the reduced density matrix is obtained as $\rho_1(t)=\operatorname{Tr}_2[\rho(t)]$.

Due to the distinct spectral topologies of this three-level system, We study the entropy evolution in three different parameter regimes ($\Omega_1<\Omega_p$, $\Omega_1=\Omega_p$, $\Omega_1>\Omega_p$). For $\Omega_1<\Omega_p$, there are two close 2nd order EP around $\Omega_2$ at 1 GHz. For one 2nd order EP, the entanglement evolution configuration is like a crescent-shaped pattern with maximum entanglement in the center near EP\cite{li2023speeding}. However, for two adjacent 2nd-order EPs, they generate an interference pattern in the intermediate parameter regime, potentially highlighting the dynamical interaction between EPs as shown in Fig. \ref{osp}. Fidelity or concurrence mapping serves as effective methods for finding this interaction region. At critical point when $\Omega_1=\Omega_p$, two EPs merge into one 3rd order EP, the entropy also collapses into a single localized peak. Notably, the time to reach its first maximum point is about $t_m\approx15$ ns--drastically faster than Hermitian situation for the same model in Appendix. \ref{appendix c}, which requires about 250 ns. When $J$ gets smaller, it shifts the $\Omega_2$ maximum entropy point closer to the EP. To quantify the effect of enhancement for the speed of entanglement generation, we introduce the ratio of the first maximum entropy generation time in the Hermitian system ($\sim J^{-1}$) and the NH system $t_m$ as a factor of enhancement. The factor of enhancement scales as $J^{-4/5}$, suggesting an effective 5th-order EP system (Fig. \ref{enhance}). This behavior stems from $n$th-order EP dynamics, where the eigenvalue splitting scales as $\Delta E \sim J^{1/n}$ \cite{hodaei2017enhanced} and the time scale of state evolution scales as $t_m \sim J^{-1/n}$. Thus, the ratio yields an enhancement of $J^{-(n-1)/n}$, which is $J^{-4/5}$ for $n=5$. Furthermore, because this entropy peak is highly localized, it is particularly sensitive to timing resolution and noise, making the system promising for high-precision quantum sensing. Interestingly, for $\Omega_1>\Omega_p$, the 3rd order EP have already disappeared, but residual EP enhancement effects persist as shown in Fig. \ref{olp}. System dynamics near this regime are similar to those at the 3rd-order EP, but with a smaller peak area and a longer generation time. This confirms that the accelerated entanglement remains robust against variations in $\Omega_1$. However, due to the broken of the chiral symmetry, the system cannot reach the maximally entangled state which has the entanglement entropy of $log_2(3)$, which also holds for the detuned Hermitian counterpart.

\begin{figure}[t]
\centering
\includegraphics[width=0.4\textwidth]{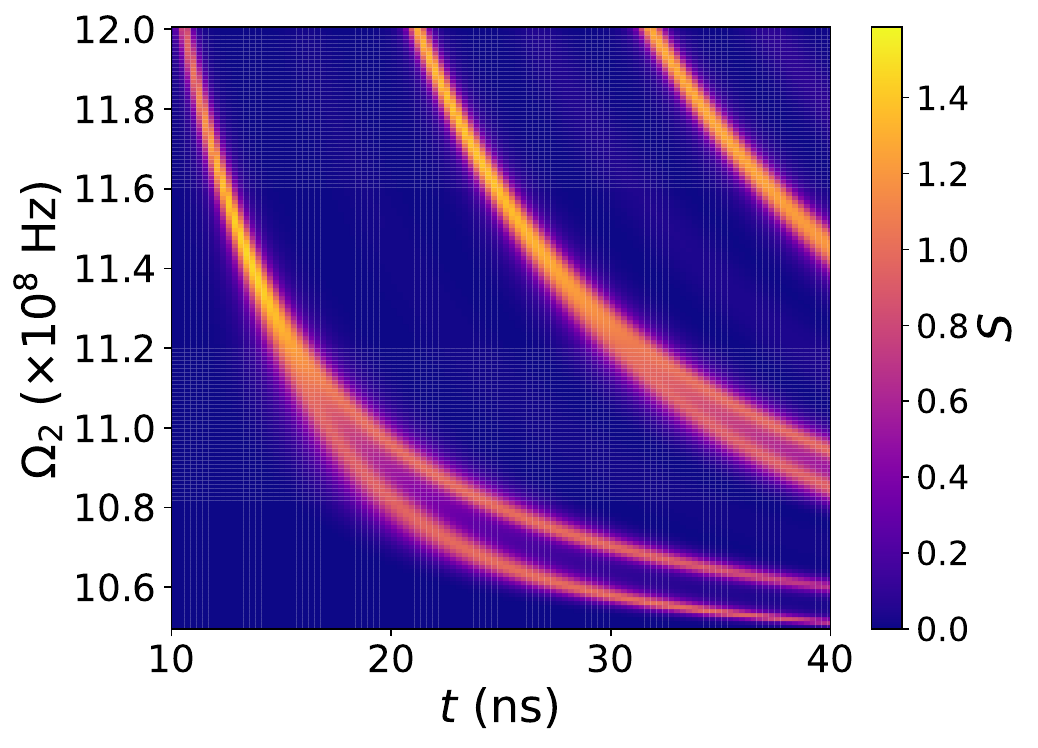}
\caption{\label{rese} Resonant NH three level system entanglement entropy evolution. The entanglement could reach near-maximum. Here we set $w_{A}=w_{B}=w_{C}=50\times10^8, e=10\times10^8$ and $J=0.1\times10^8$ (Hz).}
\end{figure}

To demonstrate that such system can also approach max entangled state and still preserves the speedup effect, we restore the effective chiral symmetry and make all three levels resonant. With $\sigma_z = 0$, the spectrum has a straightforward topological structure. To approach the maximally entangled state, both $\Omega_1$ and $\Omega_2$ should be tuned; in our model, the optimal values are $\Omega_1\approx 0.3GHz$ and $\Omega_2\approx 1.15GHz$. Despite the tight requirement, the NH speedup over the Hermitian case remains. Furthermore, to fully achieve the maximally entangled state, the two subsystems need to couple symmetrically across all subsystem combinations ($\ket{A} - \ket{B}$, $\ket{A} - \ket{C}$ and $\ket{B} - \ket{C}$) rather than only couple one pair.

In conclusion, our NH three level coupled system remains entanglement generation accelerate effect, and it could significantly approach the max entangled state under resonance. For detuned situations, the system's diverse topological structures open promising possibilities for practical quantum technologies.\\

\section{Experimental Implementation}
Our scheme can be experimentally implemented using a reconfigurable NH gauged laser array \cite{gao2023two}.
By introducing a coupling arm between two ring resonators, a propagation phase is imposed on the probability wave function during the coupling process. By precisely controlling the arm's optical path length, the evolved phase can be prepared exactly at $\pi/2$ under specific conditions. At this point, the wave function effectively experiences an imaginary coupling, thereby enabling the coupling of subsystems cavity B and cavity C.
Furthermore, the application of gain or dissipation in the coupling arm can break the symmetry of the coupling amplitude in both directions. This directional asymmetry enables asymmetric coupling between cavity A and cavity B. The NH coefficient $e$ and coupling strength $\Omega_2$ can be tuned through the gain and dissipation rate applied on coupling arms. These parameters individually controlled by a spatial light modulator (SLM), which precisely shapes the spatial profile of the optical pump beam, allowing for fine-tuned manipulation of light intensity across different arms \cite{gao2023two}. 
The cavities considered in this model are class-B semiconductor micro-ring lasers. The resonant frequencies of these laser cavities correspond to $w_i$ in mathematical model, which are $w_{A}=5\times10^9$ Hz, $w_{B}=6\times10^9$ Hz, and $w_{C}=7\times10^9$ Hz, respectively. The coupling coefficient from cavity $j$ to $i$ is denoted as $t_{i\leftarrow j}$. Specifically, $t_{B\leftarrow C}$ and $t_{C\leftarrow B}$ correspond to $i\Omega_1$, while $t_{A\leftarrow B}$ and $t_{B\leftarrow A}$ correspond to $\Omega_2\pm e$. The parameters in the model are chosen with the following logic: The imaginary part of the coupling coefficient $t$ is estimated from the threshold reduction compared to uncoupled individual lasers, the real part of $t$ is treated as a free parameter which can be utilized to adjust the magnitude of coupling \cite{dave2020complex}.
 
Nevertheless, the aforementioned setup is only valid under the condition of conservation of the excitation number, while the Hamiltonian of the system is valid within the vector space $\{\ket{1,0,0},\ket{0,1,0},\ket{0,0,1}\}$. However, due to the interaction with the environment, the quantum state can still decay to the ground state and jump outside this vector space. Thus, we need to postselect the output states and discard the ground state $\ket{0,0,0}$. With this correction of distortion state, we could obtain the Hamiltonian with dissipation $\gamma$ which can be recognized as

\begin{gather}
\begin{split}
\label{eqd}
H&=\sum_i(w_i-\frac{i}{2}\gamma_i)a_i^\dagger a_i+i\Omega_1(a_B^\dagger a_C+a_C^\dagger a_B)\\
&+(\Omega_2-e)a_A^\dagger a_B+(\Omega_2+e)a_B^\dagger a_A,
\end{split}
\end{gather}

The three laser cavities are configured to hold a similar dissipation rate ($\Delta\gamma/\gamma\ll 1$). The system will gradually decay to the vacuum state with the non-normalized concurrence approaching zero. Thus, to yield the states in the single-exciton subspace efficiently, it is necessary to perform measurements in a short time (i.e., in $1/\gamma$). On the other hand, due to the fluctuation of environment and the system itself, $\Delta\gamma$ can still affect the experimental results as calculated in Appendix \ref{appendix c}. To guarantee the measurement fidelity and minimize steady-state concurrence errors, the dissipation rates of all three cavities should be aligned to equalize their photon lifetimes and stabilize the system. In practice, the system could be first initialized in state $\ket{B}$ to simulate the single-exciton case. Through the adjustment of gain-loss rate and length of coupling arm, we can perform the above theoretical discussion on the laser array. Prepare a large number of such states, detecting the distribution of photon in three laser cavities within the photon lifetime, and discard decayed results to depict the steady state concurrence of the system. Additionally, the theoretical model of $\text{EP}_3$ can also be implemented in the flying atom regime \cite{hao2023topological}.

\section{Conclusion and discussion}
In summery, we have demonstrated a pure quantum optical system with an $\text{EP}_3$ that exhibits entanglement behavior and characteristic related to dynamical phase transition. Furthermore, with fidelity methods, We have shown that an $\text{EP}_3$ can physically arise from the merging of two $\text{EP}_2$, representing the phase transition between two distinct symmetry-broken phases. By introducing concurrence as a key entanglement measure, we investigated the NH entanglement properties of $\text{EP}_3$ and revealed that the third-order EP phase transition can be characterized by steady state concurrence, which changes sharply near the transition points. Notably, the third-order EP phase transition is recognized as a continuous phase transition. To explore the thermodynamical characteristic, we mapped out the phase diagram of the system, in which different kinds of phase transitions and their positions in the parameter space are illustrated. Additionally, we studied the dynamics of weakly coupled systems. Results show that a higher-order EP system revealing different characteristics of entanglement generation than a coupled second-order EP system, while maintaining accelerated speeds effect. Crucially, these distinct traits align with the spectrum of a single third order system. For experimental implementation, the system can be designed with a two-dimensional reconfigurable NH gauged laser array, where coupling arms connecting target cavities can be adjusted to realize the non-reciprocal interaction. Our research has deeply excavated the new properties of pure quantum optical systems and unveiled the physical connotation of high-order EPs and their potential roles in dynamic phase transitions. This work has opened a new window for future study of the application of high-order EPs and the dynamic phase transitions of quantum systems.

\begin{acknowledgments}
    
\end{acknowledgments}

\appendix

\section{Entanglement of the qubits and eigenstates}
\label{appendix a}
For a two-qubit system confined to the case of single excitation, the relevant basis states are $|0\rangle$ and $|1\rangle$. The entanglement of this two qubits can be qualified as the square root of the eigenvalues of the operator 

\begin{gather}
\label{concur}
\tilde{\rho}=\rho(\sigma^1_y\otimes\sigma^2_y)\rho^*(\sigma^1_y\otimes\sigma^2_y),
\end{gather}
where $\rho$ is the density matrix of the entangled system and $\sigma^i_y$ are the y-component Pauli operator for the $i$-th qubit, expressed as

\begin{gather}
\label{pauli}
\sigma^i_y=-i|0_i\rangle\langle 1_i|+i|1_i\rangle\langle 0_i|.
\end{gather}

Suppose that $\lambda_1\ge\lambda_2\ge\lambda_3\ge\lambda_4$ are the square roots of the eigenvalues of $\tilde{\rho}$. Then the two-qubit entanglement associated with the density matrix $\rho$ is quantified by the quantity

\begin{gather}
\label{lambda}
C=Max\{\lambda_1-\lambda_2-\lambda_3-\lambda_4, 0\}.
\end{gather}

To begin with, we consider a subsystem composed of any two qubits within the three-cavities model and analyze its entanglement properties. Since all three two-qubit subsystems exhibit similar characteristics at the phase transition, we limit our analysis to the $q_2-q_3$ subsystem for simplicity. Thus, the reduced density matrix $\rho_{BC}$ for cavity B and cavity C is presented as 

\begin{gather}
\begin{split}
\label{rd}
\rho_{BC}=&Tr_A(\ket{\psi_{all}}\bra{\psi_{all}})\\
=&\langle 0_A\ket{\psi_{all}}\bra{\psi_{all}}0_A\rangle+\langle 1_A\ket{\psi_{all}}\bra{\psi_{all}}1_A\rangle,
\end{split}
\end{gather}
here $\psi_{all}$ denotes the entangled states at a given moment $t$ expressed as $\alpha(t) \ket{1,0,0}+\beta(t)\ket{0,1,0}+\gamma(t)\ket{0,0,1}$ or, briefly, $\alpha \ket{A}+\beta\ket{B}+\gamma\ket{C}$. The result of reduced matrix in the basis $\{\ket{0_B,0_c}, \ket{0_B,1_c}, \ket{1_B,0_c}, \ket{1_B,1_c}\}$ with normalization coefficient $N$ is as follows

\begin{gather}
\label{rdf}
\rho_{BC}=N 
\begin{pmatrix}
\alpha^2&0&0&0\\
0&\beta^2&\beta\gamma^*&0\\
0&\beta^*\gamma&\gamma^2&0\\
0&0&0&0
\end{pmatrix}.
\end{gather}

The corresponding eigenvalues of operator $\tilde{\rho}$ is calculated to be $4N^2\beta^2\gamma^2$. The resulting two qubits concurrence for system states at time $t$ is then given by

\begin{gather}
\label{cnt}
C_{BC}(t)=C(t)=\frac{2|\beta(t)||\gamma(t)|}{\sqrt{\alpha(t)^2+\beta(t)^2+\gamma(t)^2}}.
\end{gather}

\section{Steady state entanglement}
\label{appendix b}
\begin{figure*}[t]
    \begin{center}
        \subfigure[\;]{\includegraphics[width=0.3\textwidth]{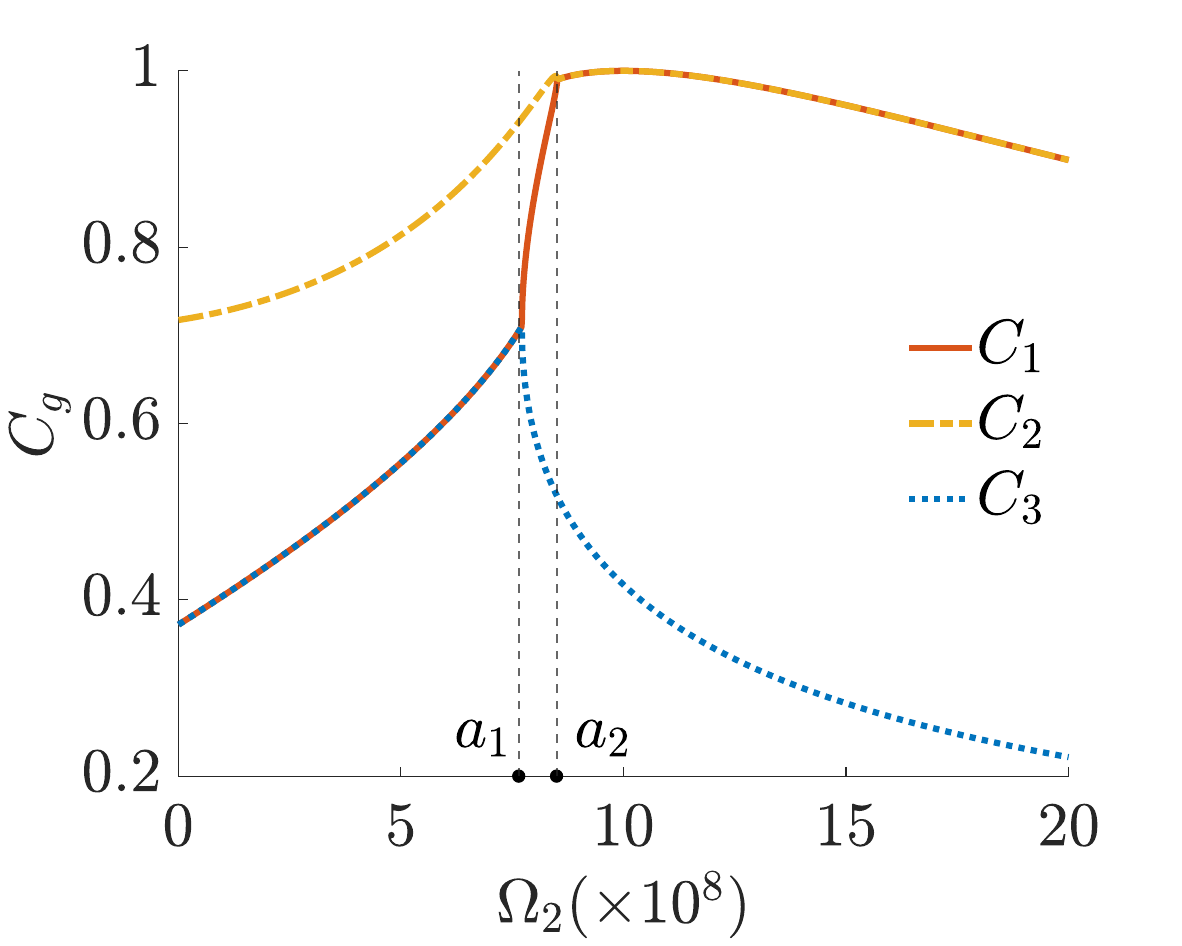}
        \label{con6}}
        \subfigure[\;]{\includegraphics[width=0.3\textwidth]{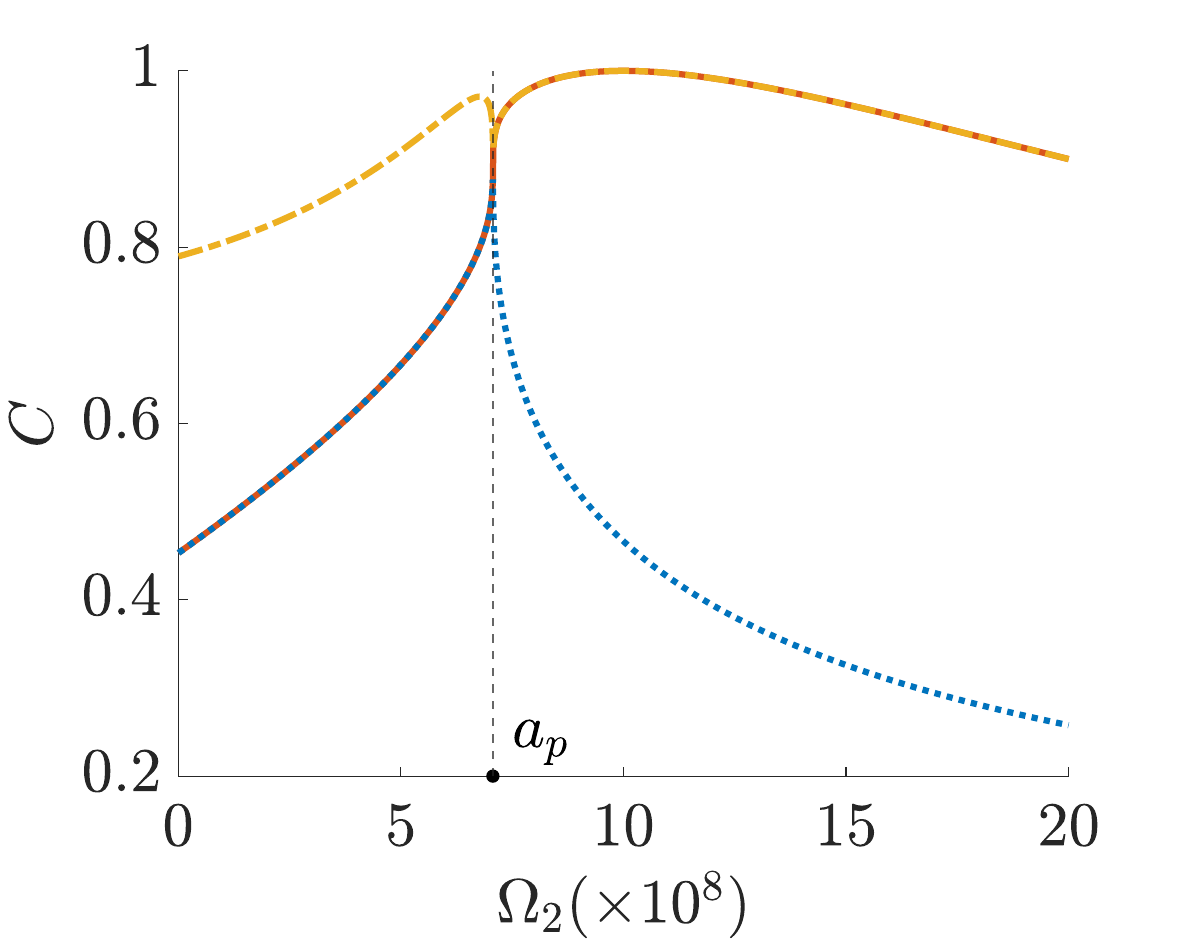}
        \label{conq}}
         \subfigure[\;]{\includegraphics[width=0.3\textwidth]{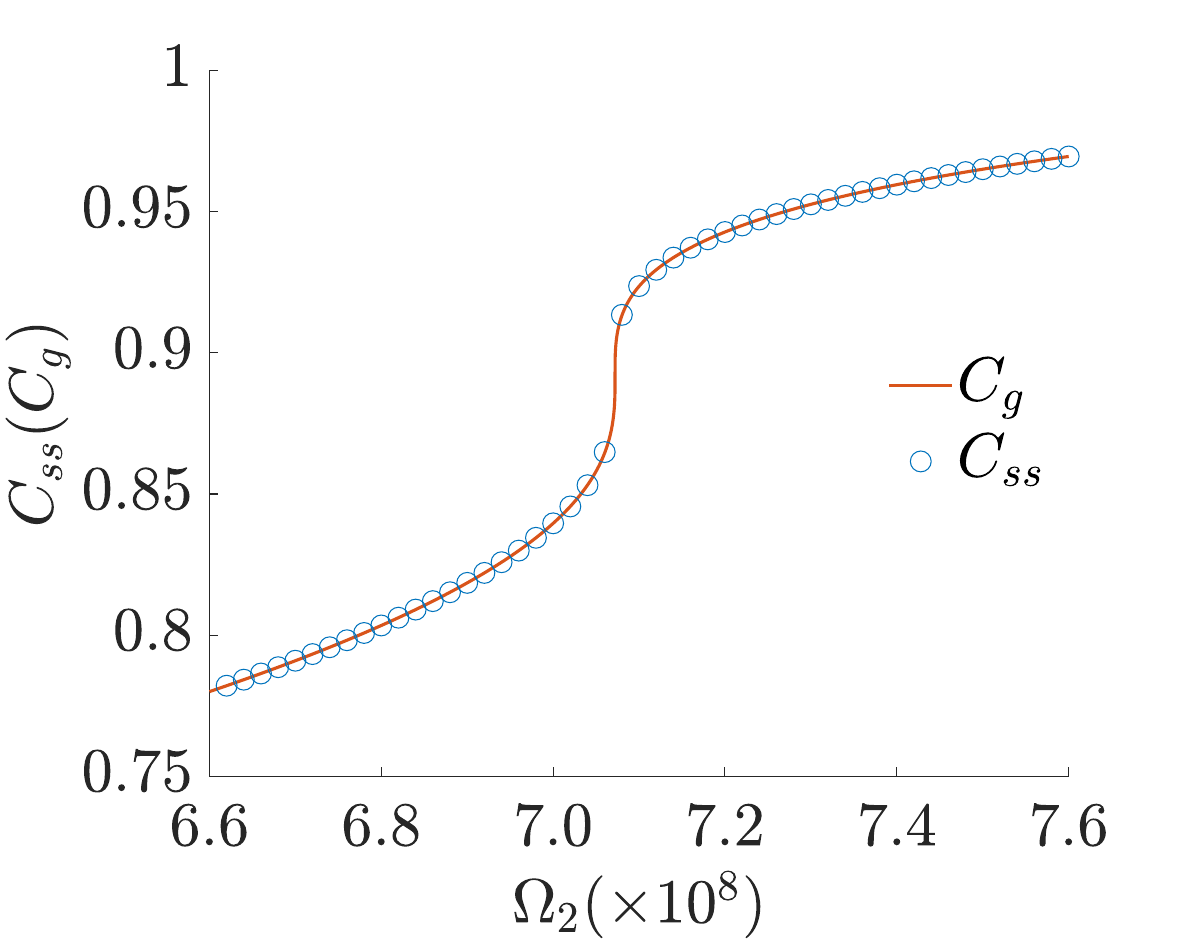}
        \label{conss}}
        \caption{\label{fig:con}Entanglements for different eigenstates and its relation with steady state concurrence. (a) Concurrences of different eigenstates with $\Omega_1=6\times10^8$ Hz. In the region $(a_1, a_2)$ all three concurrences exist without coalescing. (b) Entanglements with $\Omega_1=\Omega_p$, a high-order phase transition point emerge at $\Omega_p$. (c) $C_{ss}$ (blue spots) calculated from dynamical evolution fits well with $C_g$ curve (orange line).}
    \end{center}
\end{figure*}

For a three-level system, there are three concurrences $C_1, C_2, C_3$ corresponding to all three eigenstates $E_1, E_2, E_3$ as shown in Fig. \ref{fig:con}. It is illustrated that the eigenstate concurrences has the similar structure as the eigenenergy. When $\Omega_1<\Omega_q$ (Fig. \ref{con6}), three concurrences in the region between $a_1$ and $a_2$ are diverging separately, which corresponds to the Rabi oscillation dynamical phase. For second-order degeneracy, the phase transition at the EP can be characterized by the non-zero jump of derivatives $dC_i/d\Omega_2$ at the phase transition point \cite{han2023exceptional}. However, for an $\text{EP}_3$, it is inappropriate to characterize the phase transition by a single eigenstate concurrence $C_i$, because the original eigenstate corresponding to the gain mode will switch to another eigenstate when crossing the phase transition. Instead, it is more suitable to use the steady-state concurrence to indicate the phase transition. 

The main characteristic (such as concurrence) of the system's steady state in different phase should be determined by different eigenstates. Specifically, the gain state concurrence is given by $C_1$ for $\Omega_2<a_p$ and $C_2$ for $\Omega_2>a_p$, as shown in Fig. \ref{conq}, which is denoted as $C_g$. The reason is that when 3rd-order EP occurs ($\Omega_1=\Omega_p$), the system is in symmetry-broken phase with imaginary eigenenergies throughout $\Omega_2 < a_3$. And the eigenstate with larger positive imaginary eigenenergy (yellow line in Fig. \ref{6b}) is amplified during the dynamical evolution, while those with negative or null imaginary parts gradually dissipate. To prove that the steady state concurrence equals to the gain eigenstate concurrence,  we first represent the evolved n-level system states in a given time $t$ in the terms of eigenstates as ($\hbar=1$)

\begin{gather}
\label{ges}
\psi(t)=\sum_i^nc_ie^{-iE_it}\ket{i},
\end{gather}
where $E_i$ are the eigenvalues of corresponding eigenstates $\ket{i}$ and $c_i$ are the expansion coefficients. For a NH system, $E_i$ may be complex in spectral decomposition. As the system reaches a steady state (i.e., $t\rightarrow\infty$), the state of the system tends to be $\sum_k e^{E_gt}\ket{k}$, where $E_g$ denotes the eigenvalue with the maximum imaginary part among all eigenvalues, and $\ket{k}$ are the corresponding degenerate eigenstates. Thus, the steady state is exactly the gain eigenstates, which implies the equality of their concurrences

\begin{gather}
\label{gcsc}
C_{ss}=C_{g}.
\end{gather}

We can further verify this deduction by calculating the steady-state concurrence of the system and comparing it with $C_g$.  The $C_{ss}$ for different values of $\Omega_2$ at $\Omega_1 = \Omega_p$ is shown in Fig. \ref{conss}. The statistical analysis demonstrates that $C_g \approx C_{ss}$.

\section{Entanglement generation in Hermitian system}
\label{appendix c}

To compare the entanglement generation speed between non-Hermitian system and Hermitian system, we reduce the Hamiltonian of Eq. \ref{eqht} to be in Hermitian form

\begin{align}
\label{hermi}
H&=\sum_iw_ia_i^\dagger a_i+\Omega_1(a_B^\dagger a_C+a_C^\dagger a_B)+\Omega_2(a_A^\dagger a_B+a_B^\dagger a_A).
\end{align}

\begin{figure}[t]
    \begin{center}
        \subfigure[\;detune]{\includegraphics[width=0.23\textwidth]{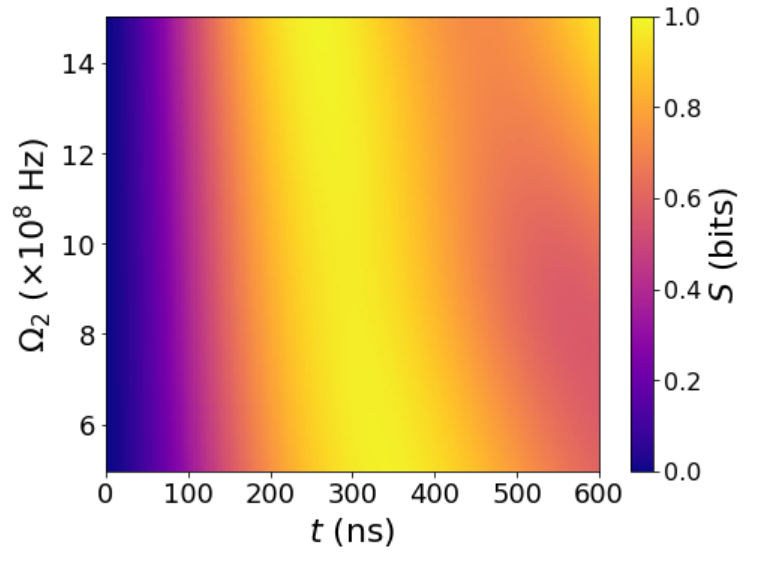}
        \label{deher}}
        \subfigure[\;resonant]{\includegraphics[width=0.23\textwidth]{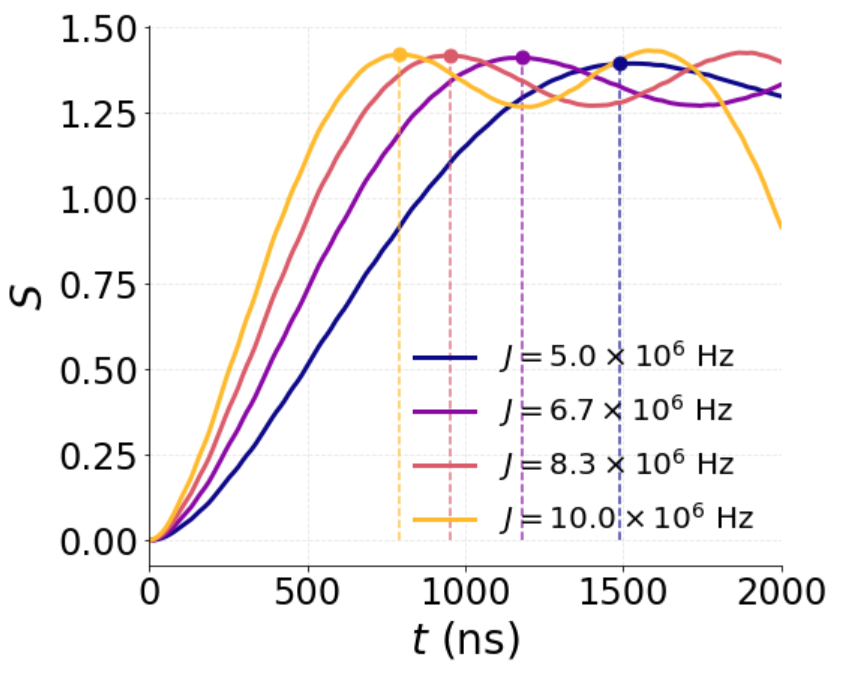}
        \label{reher}}
        \caption{\label{fig:cone}(a) Entanglement entropy evolution as a function of $\Omega_2$ in detuned Hermitian system. $w_{A}=50\times10^8, w_{B}=60\times10^8, w_{C}=70\times10^8$, $\Omega_1=7.07\times10^8$ and $J=0.1\times10^8$ (Hz). (b) Entanglement entropy evolution for certain $\Omega_2=3\times10^8$ and $\Omega_1=3\times10^8$ where $w_{A}=w_{B}=w_{C}=50\times10^8$ (Hz).}
    \end{center}
\end{figure}

To compare with the NH case of 3rd EP, here we also set absolute value of $\Omega_1=\Omega_p$. The result shows that the first maximum entanglement is generated at $t\approx 250ns$ as shown in \ref{deher}. And since it is detuned, the max entanglement entropy is also near only 1. Similarly for resonant case, we set all $w_i=5 GHz$ and $\Omega_1=0.3 GHz$. As depicted in \ref{reher}, the results show the first maximum entropy will be around 750 ns (NH is around 15ns), and Hermitian system will exhibit $J^{-1}$ entanglement generation time limit. To make system generate maximum entangled state with entropy equal to $log_2(3)$, we need to add up another two coupling with same coupling strength:

\begin{align}
\label{hermic}
H = \sum_{j=1,2} H_j + J\boldsymbol{\sigma}_+^{(1)} \boldsymbol{\sigma}_-^{(2)} + \text{h.c.}
\end{align}
where $\boldsymbol{\sigma}_+^{(j)} = \begin{pmatrix} \sigma_{+, AB}^{(j)} & \sigma_{+, BC}^{(j)} & \sigma_{+, CA}^{(j)} \end{pmatrix}$, $\boldsymbol{\sigma}_-=\boldsymbol{\sigma}_+^\dagger$ represents the transition operator vector for subsystem $j$. In this case, both Hermitian and our NH model could reach max entangled state while keeps EP induced entanglement acceleration.

\section{Decay situation}
\label{appendix D}
The cavities considered in this model are class-B semiconductor microring lasers. By means of postselection, we assume that the total excitation in the system is conserved to be one. If the magnetic disorder is sufficiently small to be neglected, the rate equations can be approximated as \cite{gao2023two}

\begin{align}
\begin{split}
\label{eqc}
\frac{dY_i}{dt}&=-\frac{1}{2\tau_{pi}}Y_j-\sum_jIm[t_{i\leftarrow j}e^{i(\phi_j-\phi_i)}]Y_j\\
\frac{d\phi_i}{dt}&=-\Omega_i-\frac{\alpha_H}{2\tau_{pi}}+\sum_j\frac{Y_j}{Y_i}Re[t_{i\leftarrow j}e^{i(\phi_j-\phi_i)}]Y_j.
\end{split}
\end{align}
where $Y_i$ and $\phi_i$ are the dimensionless electric field magnitudes and phases for the cavity lasing modes, $\Omega$ is the resonant frequency, parameter $\alpha_H=2$ and, $\tau_{pi}$ represent photon lifetimes for different cavities. The coupling coefficient from cavity $j$ to $i$ is denoted as $t_{i\leftarrow j}$, and the index $j$ sums over the nearest neighbors. To match the theoretical model above, the electric field magnitudes can be regarded as amplitude magnitudes of different basis states. $\Omega_i$ are set to be $w_i$. $t_{2\leftarrow 3}$ and $t_{3\leftarrow 2}$ equal to $i\Omega_2$. $t_{1\leftarrow 2}(t_{2\leftarrow 1})$ equals to $\Omega_1\pm e$. When $\tau_{p1}=\tau_{p2}=\tau_{p3}=\tau_{0}$ ($\tau_{0}=2\times10^{-8}$ s), the rate equations will lead to the same result as the decay-free model. While the nonnormalized concurrence will indicate the decay rate of the amplitude or concurrence.

\begin{figure}[t]
    \begin{minipage}[b]{0.48\columnwidth}
        \centering
        \subfigure[]{\includegraphics[width=1.1\textwidth]{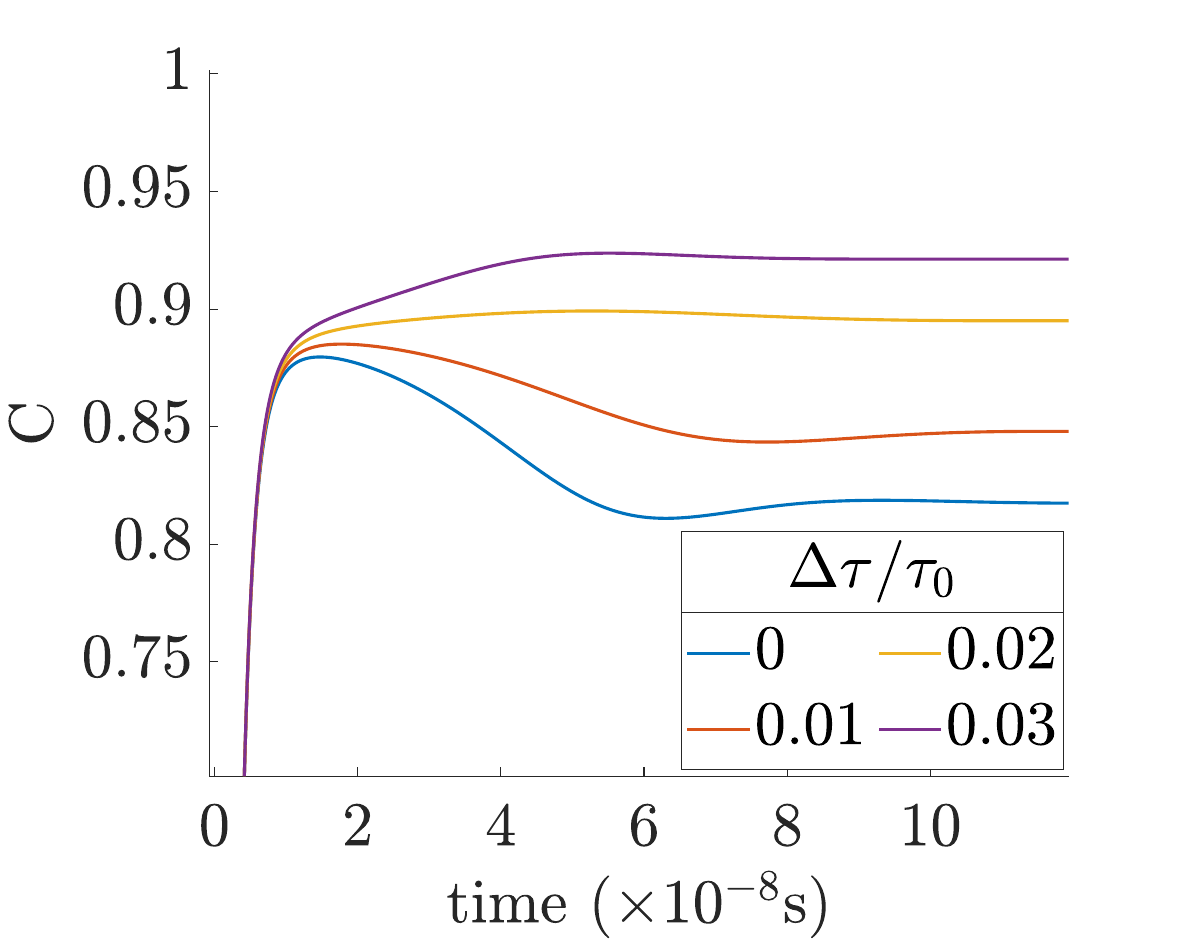}
        \label{dconss}}
    \end{minipage}
    \begin{minipage}[b]{0.48\columnwidth}
        \subfigure[]{\includegraphics[width=1.1\textwidth]{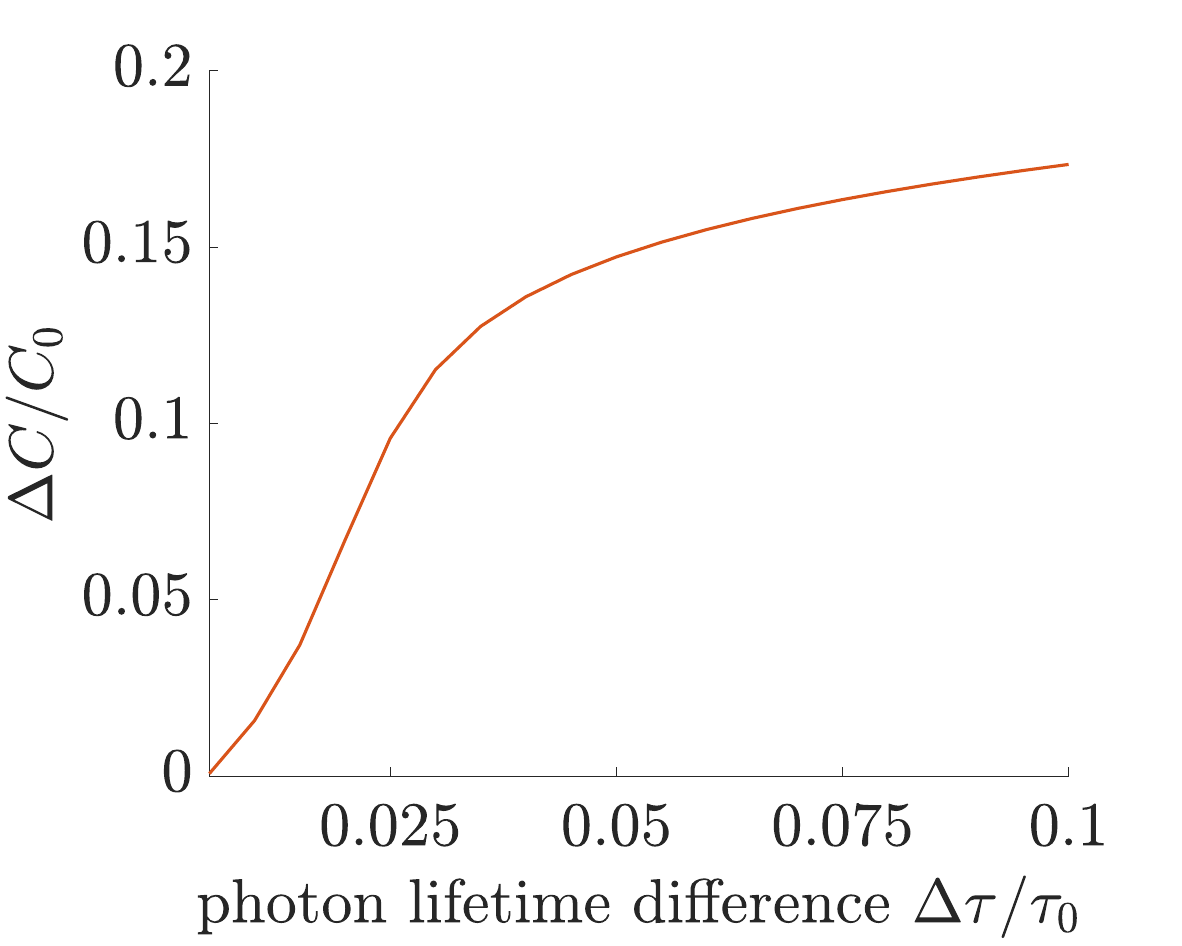}
        \label{decayr}}
    \end{minipage}
        \caption{\label{fig:laser}Quantitative decay error analysis. (a) The influence of photon lifetime uncertainty on steady state concurrence. (b) The error of steady state concurrence respect to decay rate difference.}
\end{figure}

When one of cavities photon lifetime changes due to environmental disturbances, i.e. $\tau_{pi}-\tau_0=\Delta\tau$, the steady state will shift accordingly as shown in Fig. \ref{dconss}. The blue curve at the bottom represents the evolution curve with $\Delta\tau=0$ where all three cavities' photon lifetimes align well. As $\Delta\tau$ increases, the concurrence time evolution will increasingly deviate from the original standard one. The ratio of steady-state concurrence deviation $\Delta C/C_0$ and photon lifetime error $\Delta\tau/\tau_0$ has been configured in Fig. \ref{decayr}.

\bibliographystyle{apsrev4-2} 
\bibliography{ref}

\end{document}